\documentclass{article}

\usepackage{PRIMEarxiv}

\DeclareFontFamily{U}{mathx}{\hyphenchar\font45}
\DeclareFontShape{U}{mathx}{m}{n}{
      <5> <6> <7> <8> <9> <10>
      <10.95> <12> <14.4> <17.28> <20.74> <24.88>
      mathx10
      }{}
\DeclareSymbolFont{mathx}{U}{mathx}{m}{n}
\DeclareMathSymbol{\bigtimes}{1}{mathx}{"91}

\usepackage{bm}
\usepackage{url}
\usepackage{amsmath}
\usepackage{amsfonts}
\usepackage{amssymb}
\usepackage{amsthm}
\usepackage{bbm}
\usepackage{subcaption}
\usepackage{diagbox}
\usepackage{multirow, makecell}
\usepackage[utf8]{inputenc} 
\usepackage[T1]{fontenc}      
\usepackage{url}            
\usepackage{booktabs}              
\usepackage{nicefrac}       
\usepackage{microtype}      
\usepackage{lipsum}
\usepackage{fancyhdr}       
\usepackage{graphicx}       
\usepackage[round]{natbib}
\graphicspath{{media/}}     

\usepackage{ulem}
\usepackage{collcell} 
\usepackage{pgf}

\definecolor{royalblue}{RGB}{65, 105, 225} 
\definecolor{darkorange}{RGB}{255, 140, 0}  
\definecolor{darkgreen}{RGB}{0, 150, 0}

\newtheorem{lemma}{Lemma}

\usepackage{moreverb,url}

\usepackage[colorlinks,bookmarksopen,bookmarksnumbered,citecolor=red,urlcolor=red]{hyperref}
\usepackage{booktabs}

\newcommand\BibTeX{{\rmfamily B\kern-.05em \textsc{i\kern-.025em b}\kern-.08em
T\kern-.1667em\lower.7ex\hbox{E}\kern-.125emX}}

\usepackage{bm}
\usepackage{xcolor}
\def\mydefb#1{\expandafter\def\csname b#1\endcsname{\bm{#1}}}
\def\mydefallb#1{\ifx#1\mydefallb\else\mydefb#1\expandafter\mydefallb\fi}
\mydefallb ABCDEFGHIJKLMNOPQRSTUVWXYZabcdeghijklnopqrstuvwxyz\mydefallb

\def\mydefgreek#1{\expandafter\def\csname b#1\endcsname{\text{\boldmath$\mathbf{\csname #1\endcsname}$}}}
\def\mydefallgreek#1{\ifx\mydefallgreek#1\else\mydefgreek{#1}%
	\lowercase{\mydefgreek{#1}}\expandafter\mydefallgreek\fi}
\mydefallgreek {beta}{Gamma}{alpha}{Delta}{epsilon}{Omega}{Theta}{Iota}{Lambda}{kappa}{mu}{nu}{Xi}{Pi}{chi}{rho}{Sigma}{Psi}{tau}\mydefallgreek
\newcommand{\C}{\text{\normalfont C}}
\newcommand{\D}{\text{\normalfont D}}
\newcommand{\Iend}{{\bar{i}}}
\usepackage{accents}
\usepackage{placeins}
\newcommand{\ubar}[1]{\underaccent{\bar}{#1}}
\newcommand{\Tau}{{\text{T}}}
\newcommand{\bff}{{\bm f}}

\newtheorem{definition}{Definition}
\newtheorem{remark}{Remark}
\usepackage[toc,page]{appendix}

\title{A burn-in(g) question: How long should an initial equal randomization stage be before Bayesian response-adaptive randomization?}

\author{
  Edwin Y.N. Tang$^\dagger$, Stef Baas$^{\dagger}$, Daniel Kaddaj, 
Lukas Pin,  David S. Robertson, and Sofía S. Villar \\ \\
  MRC Biostatistics Unit \\
  University of Cambridge \\
  CB2 0SR \\
  Cambridge\\
  $^{\dagger}$: First authors
}

\begin{document}
\maketitle
\begin{abstract}
Response-adaptive randomization (RAR) can increase participant benefit in clinical trials, but also complicates statistical analysis. The burn-in period—a non-adaptive initial stage—is commonly used to mitigate this disadvantage, yet guidance on its optimal duration is scarce. To address this critical gap, this paper introduces an exact evaluation approach to  investigate how the burn-in length impacts statistical operating characteristics of two-arm binary Bayesian RAR (BRAR) designs.
		We show that (1) commonly used calibration and asymptotic tests show substantial type I error rate inflation for BRAR designs without a burn-in period, and increasing the total burn-in length to more than half the trial size reduces but does not fully mitigate type I error rate inflation, necessitating exact tests; (2) exact tests conditioning on total successes show the highest average and minimum power up to large burn-in lengths; (3) the burn-in length substantially influences power and participant benefit, which are often not maximized at the maximum or minimum possible burn-in length;  (4) the test statistic influences the type I error rate and power; (5) estimation bias decreases quicker in the burn-in length for larger treatment effects and increases for larger trial sizes under the same burn-in length. Our approach is illustrated by re-designing the ARREST trial.
\end{abstract}


\maketitle
\section{Introduction}\label{sect:intro}
Clinical trials are important studies that
	evaluate the effects of new treatments on human health outcomes.Randomization is typically used in confirmatory clinical trials (and recommended where possible in Phase~II settings) because it induces comparable treatment groups, mitigates selection bias and can provide a basis for statistical inference~\citep{rosenburger2016rand}.
	More often than not, modern-day clinical trials use a fixed
	randomization scheme~(usually the permuted block design). Alternatively, statisticians can consider response-adaptive randomization~(RAR), which allows allocation probabilities to change based on previous allocations and outcomes. 
	A well-designed RAR procedure typically aims to balance the 
	goal of drawing correct inferential conclusions with that  of maximizing participant benefit. 
	
	Response-adaptive (RA) procedures have been used in exploratory or seamless phase II/III multi-arm trials~\citep[see, e.g.,][]{Berry2023comment}.
	Furthermore, a publicly available list
	summarizing RA clinical trials in the last 100 years currently reports 10 out of 30  RA trials were confirmatory~\citep{Pin2025}.
	Despite this, the fraction of confirmatory trials with an RA component remains relatively low.
	A reason for this could be the ongoing debates about the risks associated with RA designs~\citep[see, e.g.,][for an overview]{Robertson_Lee_López-Kolkovska_Villar_2023}. 
	While arguments in favour emphasize the prospect of improving participant  
	benefit~\citep{Rosenberger_Sverdlov_Hu_2012}, counterarguments highlight the non-negligible possibility of assigning more participants to the inferior arm~\citep{thall2015statistical}. The risk of substantial between-arm imbalances is especially worrisome when there is accrual bias~(e.g., more severely ill patients being enrolled earlier on) or a temporal trend in prognostic baseline characteristics~\citep{Proschan2020Resist}. The use of an RA design also impacts the statistical analysis, since classical statistical methods may not
	maintain their desirable and well-understood properties such as type I error rate control~\citep{baas2025exact}.
	
	A common and ad hoc approach to alleviate the weaknesses mentioned above is the inclusion of a period of non-adaptive (fixed) allocation at the start of the RA design, which we will refer to from now on as a {\it burn-in period}. 
	Computational results in \citet{du2018comparing} show that in comparison to a fixed non-response-adaptive design with equal randomization, a suitable burn-in period length for a {Bayesian~RAR}~(BRAR) design allows more participants to be assigned to the superior arm on average due to some response-adaptiveness, but at the cost of a small decrease in statistical power.
	
	Although the importance of a burn-in phase is generally recognized, few studies offer a rationale or provide practical guidance on its duration.  The absence of a robust justification for burn-in period length is a notable deficiency in the current literature. \citet{Robertson_Lee_López-Kolkovska_Villar_2023} only touch upon burn-in in their review, and \citet{thorlund2018key}'s recommendation of 20--30 patients per arm may be overly broad, as the optimal length likely varies with the RA procedure, primary outcome type and sample size.
	\citet{viele2020comparison2, viele2020comparison} considered the effect of burn-in length for multi-arm RA designs, while~\citet{GRANHOLM202345} considered general adaptive designs.
	Our approach differs from theirs in that we focus on type~I error rate control across the null parameter set and numerically evaluate multiple metrics to inform burn-in recommendations for two-arm BRAR designs. 
	As demonstrated in Table~\ref{sect:BRARtab}, burn-in periods are common in BRAR trials, yet the existing reports rarely include a transparent explanation or rationale for the chosen duration.
	
	To address the above gap, we concentrate on two-arm BRAR designs using a burn-in period as the sole tuning parameter and which use the posterior probability of control superiority to test for a treatment effect. The BRAR design is prevalent in implemented RA trials~\citep{Pin2025}, whereas this test statistic is prevalent in BRAR designs using a burn-in~(Table~\ref{sect:BRARtab}). To ensure a focused evaluation, we treat the burn-in period as the exclusive mechanism of regularization, thus isolating its influence on operating characteristics from other design adjustments~(such as clipping or tuning). Importantly, our exact analysis framework is generalizable to different test statistics and to other BRAR variations, such as batched allocation, clipping, and power transformations~\citep[e.g.,][]{du2018comparing}.
	In this context our contributions are as follows, we (1)~exactly assess the effect of the burn-in length on the type I error rate, power, participant benefit, and the probability of an imbalance in the wrong direction in a BRAR clinical trial design. 
	(2)~We compute the point-wise, average, minimum, and maximum operating characteristics over the parameter space, where the last three measures summarize the dependence of operating characteristics on the burn-in length over the complete parameter space. Our approach 
	avoids Monte Carlo error~\citep{koehler2009assesment}, which allows us to, e.g., compute the optimal burn-in length in terms of power, something that is much harder to estimate through simulation~(due to non-smoothness).
	(3)~We construct conditional and unconditional exact tests for BRAR designs with a burn-in and compare them to commonly used calibrated or asymptotic tests. Exact tests for RA designs, introduced in~\citet{wei1990statistical}, bound the type I error rate above by the target significance level for all possible parameters under the null hypothesis.
	This strong form of type I error rate control is often desired in confirmatory trials.
	Although such exact tests are not novel,  they have mostly been limited to fully sequential RA procedures in the literature, whereas our paper considers BRAR designs with a burn-in period and a group-sequential BRAR design in the real-life application.
	(4)~By default, exact inference is computationally more demanding than inference using asymptotic tests. Similar to~\citet{baas2025exact}, we use the efficiency considerations outlined in~\citet{jacko2019binarybandit} to efficiently compute design operating characteristics and exact tests.
	As a result, we are able to evaluate RA designs with all possible burn-in lengths for trial sizes of up to 240 participants, while exact approaches for RA designs in the literature are often limited to less than 100 participants.
	(5)~Finally, we use our novel findings to provide suggestions on how to choose the burn-in in clinical trials using a BRAR design.
	
	This paper is structured as follows: Section \ref{sect:model} introduces the model and notation for a two-arm RA design with a burn-in period. Section~\ref{sect:inferencemethods} introduces statistical methods. Section~\ref{typeIerrorcurves} provides a numerical investigation to assess to what extent the burn-in period can be used to control type I error rates for commonly used tests, Section~\ref{sect:exact_tests} investigates the added value of a burn-in period for BRAR designs using exact tests, Section~\ref{sect:participant} evaluates the impact of the burn-in length on participant benefit metrics, Section~\ref{sect:bias_results} evaluates the impact of the burn-in length on estimation bias, Section~\ref{sect:recommend} gives recommendations for the burn-in length, and Section~\ref{sect:sensitivity_analysis} provides a sensitivity analysis of our findings with respect to the choice of prior, where our default is the uniform prior. Section~\ref{sect:ARREST}
	applies our proposed methodology to 
	a real-world BRAR clinical trial that used blocked allocation and early stopping~(the ARREST trial). Section~\ref{sect:discussion} summarizes the findings and presents directions for future research. The appendix of the paper contain a summary table of BRAR trials with a burn-in
	period, results for the same analysis as in Section~\ref{typeIerrorcurves} and Section~\ref{sect:exact_tests} but for the Wald test instead of the posterior probability test, critical values of the exact tests of Section~\ref{sect:exact_tests}, and additional results for the ARREST trial analysis~(Section~\ref{sect:ARREST}).

\section{Two-arm response-adaptive design with a burn-in period}~\label{sect:model}
This section provides the model and notation for a two-arm response-adaptive~(RA) clinical trial with binary outcomes and a burn-in period. 
In this paper, we will mainly follow the notation of \cite{baas2025exact}. 
We consider the parametric model where~$\btheta=(\theta_\C,\theta_\D)$ are the unknown success probabilities of the control treatment~(C) and the developmental treatment~(D) respectively. In the remainder, the same ordering of the treatment indicators (i.e., first~$\C$ then~$\D$) will be used to construct vectors. Let~$\bY_\C=(Y_{\C,i})_{i=1}^\Iend$ and~$\bY_\D=(Y_{\D,i})_{i=1}^\Iend$ be two sequences of independent Bernoulli random variables, where~$\mathbb{P}_\btheta(Y_{a,i}=1)=\theta_a$ for~$a\in\{\C,\D\}$.  The random variable~$Y_{a,i}$ denotes the potential outcome under treatment~$a$ for trial participant~$i$, while the natural number~$\Iend$ denotes the fixed trial size.

In a two-arm RA clinical trial,  participants~$i\in \mathcal{I}=\{1,2,\ldots,\Iend\}$ arrive sequentially, and each participant is allocated to a treatment arm~$A_i$, resulting in a  response~$Y_{A_i,i}$. \hbox{Let~$\bH_i=(A_1,Y_{A_1,1},\ldots,A_i,Y_{A_i,i})$} be the trial history up to and including participant~$i\in\mathcal{I}$. Denote the support set of all trial histories by~$\mathcal{H}=\bigcup_{i=0}^\Iend \mathcal{H}_i$ where~$\mathcal{H}_0=\emptyset$ and 
\begin{equation*}
    \mathcal{H}_i=\{(a_1,y_1,\ldots,a_i,y_i):\;y_w\in\{0,1\},\;a_w\in\{\C,\D\}\;\; \forall w\in\{1,\ldots,i\}\}.
\end{equation*}
An RA procedure is a function~$\pi: \mathcal{H}\mapsto [0,1]$, where~$\pi(\bH_i):=\mathbb{P}^\pi(A_{i+1}=\C\mid\bH_i)$. The joint probability measure on the outcomes and allocations induced by the RA procedure will henceforth be denoted by~$\mathbb{P}_\btheta^\pi.$ With the above notation we can now define the total successes and treatment group sizes up to participant~$i$, defined respectively as:
\begin{equation*}
\textstyle
    S_{a,i}=\sum_{i'=1}^i Y_{A_{i'},i'}\mathbb{I}(A_{i'}=a), \qquad N_{a,i}=\sum_{i'=1}^i \mathbb{I}(A_{i'}=a),\quad a\in\{\C,\D\}.
\end{equation*}
Letting~$\mathcal{I}_0=\{0\}\cup\mathcal{I}$, where~$i=0$ represents the point at which no participant outcomes have been collected (i.e., the start of the trial), we  define~$\bX_i=(\bS_i,\bN_i)$ as the tuple containing the total successes and treatment group sizes with support $$\mathcal{X}_i=\{((s'_\C,s'_\D),(n'_\C,n'_\D)):\bs',\bn'\in\mathcal{I}_0^2, \bs'\leq \bn',\;\textstyle n'_\C+n'_\D=i\}.$$
The tuple~$\bX_{\Iend}$, consisting of the total successes and treatment group sizes at the end of the trial, which are the sufficient~(summary) statistics for the Bernoulli (exponential family) model, can be used to determine many estimators and test statistics~(e.g., the Wald, score and likelihood ratio test statistic, as well as the maximum likelihood estimator for~$\btheta$).

In this paper, we will consider the BRAR procedure with a burn-in length~$b$ per arm, denoted~$\pi^{\text{B}}_b$.
Under this procedure, the first~$2b$ trial participants 
for a \hbox{burn-in~$b\in\{0,\dots,\Iend/2\}$}
are allocated to treatment in an non-response-adaptive manner, such that the treatment group sizes deterministically equal~$b$ after allocating participant~$2b$, i.e.,~$N_{a,2b}=b$ for all~$a\in\{\C,\D\}.$
After the burn-in phase has been completed, i.e., participant~$i$ has to be allocated for~$i>2b$, the procedure becomes response-adaptive and participants are allocated to the control treatment with probability equal to the posterior probability that the control treatment is superior.
In the present paper, we assume the Beta(1,1) (i.e., uniform) prior, from which it follows that for~$i>2b$ we have for all~$\bx_i\in\mathcal{X}_i$, 
\begin{equation} \pi_b^{\text{B}}(\bx_i)= \int_{\theta_\C\geq \theta_\D}\prod_{a\in\{\C,\D\}}d\text{Beta}(\theta_a\,;\,s_a(\bx_i)+1,\, n_a(\bx_i)-s_a(\bx_i)+1)\,d\btheta,\label{defn:PPCS}\end{equation}
where~$d\text{Beta}$ denotes the density of the Beta distribution and the functions~$s_{a},n_a$ are defined such that~$n_{a}(\bX_i)=N_{a,i}$ and~$s_a(\bX_i)=S_{a,i}$ for~$a\in\{\C,\D\}.$

The RA procedure~$\pi^\text{B}_b$ is a~\emph{Markov RA procedure}~\citep{yi2013exact}, which means that the RA procedure can be written as a \hbox{function~$\pi:\mathcal{X}\mapsto[0,1]$} where~$\mathcal{X}=\cup_i\mathcal{X}_i$.
Under a Markov RA procedure~$\pi$,~$(\bX_i)_i$ is a Markov chain with initial state~$\bX_0=((0,0),(0,0))$, state space~$\mathcal{X},$ and transition structure
$$
    \mathbb{P}^\pi_{\btheta}(\bX_{i+1} = \bx_{i+1}\mid\bX_i=\bx_i) = \begin{cases}
\theta_\C\cdot \pi(\bx_i),\quad &\text{ if~$\bx_{i+1} = \bx_i +  \partial \bs_\C$,}\\
(1-\theta_\C)\cdot \pi(\bx_i),\quad &\text{ if~$\bx_{i+1} = \bx_i +\partial \bff_\C$,}\\
\theta_\D\cdot (1-\pi(\bx_i)),\quad &\text{ if~$\bx_{i+1} = \bx_i +  \partial \bs_\D$,}\\
(1-\theta_\D)\cdot (1-\pi(\bx_i)),\quad &\text{ if~$\bx_{i+1} = \bx_i +\partial \bff_\D$,}\\
0,&\text{ else,}
\end{cases}$$
where~$\partial\bs_\C = ((1,0),(1,0))$ and~$\partial\bff_\C = ((0,0),(1,0))$ are the change in~$\bX_i$ after a success and failure for the control arm, and~$\partial \bs_\D$,~$\partial\bff_\D$ are defined similarly.

For Markov RA procedures, it was shown in~\citet{yi2013exact} that the likelihood can be written as
\begin{equation}
    \mathbb{P}^\pi_{\btheta}(\bX_i=\bx_i)=g_i^\pi(\bx_i)\prod_{a\in\{\C,\D\}}\theta_a^{s_a(\bx_i)}(1-\theta_a)^{n_a(\bx_i)-s_a(\bx_i)}\label{eqn:likelihood}
\end{equation}
where $g_i^\pi$ represents the part of the distribution of~$\bX_i$ found by summing the probabilities of all allocation paths that lead to~$\bX_i$ and  is defined as $g_i^\pi(\bx_i)=0$ for~$\bx_i\in\mathbb{Z}^4\backslash\mathcal{X}_i$ and otherwise  recursively by
 \begin{align*}
g_0^\pi(\bx_0)&=1,\\
     \quad g_{i}^\pi(\bx_i) &= \sum_{\substack{a\in\{\C,\D\}\\\partial\bx_a\in\{\partial\bs_a,\partial{\bff}_a\}}}g_{i-1}^\pi(\bx_{i}-\partial\bx_a)\pi(\bx_{i}-\partial\bx_a)^{\mathbb{I}(a=\C)}(1-\pi(\bx_{i}-\partial\bx_a))^{\mathbb{I}(a=\D)}.
\end{align*} \label{eqn:defGcoefs}

\begin{remark}{\bf Allocation method during burn-in period.}\label{remark:burnin}
There are multiple procedures to allocate participants during the burn-in period. For small burn-in lengths in particular, it can be very important to aim for a small probability of treatment imbalances during the burn-in period. Several allocation procedures, such as the truncated binomial design, big stick design, permuted block design, and the random allocation rule can be used, where each procedure has its advantages and difficulties~\citep{berger2021roadmap}. 
In this remark, we want to emphasize that, while being a relevant topic in practice, it does not matter for our evaluation which allocation method is used during the burn-in period so long as the allocation method allocates~$b$ participants to each treatment arm. In that case, we have under the two-arm RA clinical trial model described above that:~$$g^{\pi_b^\text{B}}_{2b}(\bx_{2b})=\binom{b}{s_\D(\bx_i)}\binom{b}{s_\C(\bx_i)}\cdot\mathbb{I}(n_\C(\bx_{2b})=b).$$
Hence, assuming the outcomes are i.i.d., the specific allocation procedure used during the burn-in period does not affect the distribution of~$\bX_{\Iend}$.
\end{remark}

	\section{Statistical analysis}\label{sect:inferencemethods}
	
	In this section, we first focus on tests for the null hypothesis
	$$H_0:\theta_\D=\theta_\C\quad\text{versus}\quad H_1:\theta_\D\neq\theta_\C,$$
	after which we consider the exact calculation of trial operating characteristics.  
	In the main paper we will focus on tests for~$H_0$ that use the \emph{posterior probability of control superiority}~(PPCS), equal to the right-hand side of~\eqref{defn:PPCS} with~$i=\bar{i}$
	(i.e, equal to~$\pi_0^{\text{B}}(\bx_\Iend)$
	for all~$\bx_{\Iend}\in\mathcal{X}_{\Iend}$). 
	
	The expression of the likelihood~\eqref{eqn:likelihood} facilitates an exact analysis of the trial data, as well as the exact calculation of trial operating characteristics. 
	In the following two subsections, we first describe exact tests for testing~$H_0$ (Section~\ref{sect:exact}),
	after which we provide methods to efficiently calculate operating characteristics based on~\eqref{eqn:likelihood}~(Section~\ref{OCscalc}).
	
	\subsection{Exact tests} \label{sect:exact}
	
	\subsubsection{Conditional exact test based on total successes}\label{sect:RT}
	We first introduce the conditional test based on total successes and show that it is an exact test. The conditional test based on total successes generalizes Fisher's exact test~\citep{Fisher1934method} under a design with fixed treatment group sizes to response-adaptive 
	designs. The conditional test constructs a critical value from the conditional distribution of the test statistic given  the total sum of successes~$\text{S}(\bx_\Iend)=\sum_{a\in\{\C,\D\}}s_a(\bx_\Iend)$
	in the trial, where the nuisance parameter~$\theta = \theta_\C=\theta_\D$ under the null hypothesis is eliminated by conditioning on~S$(\bX_{\Iend})$.\\

	\begin{definition}[(Conditional test based on total successes)] \label{def:CX_SA}
		Let~$\mathcal{X}_S(s')$ be the pre-image of~$s'\in\mathcal{I}$ under~S.
		A conditional test based on~S for test statistic function~$\Tau$, significance level~$0<\alpha<1$, and  RA~procedure~$\pi$   rejects \mbox{when~$\Tau(\bX_{\Iend})\geq \bar{c}
			(\text{S}(\bX_\Iend))$
			or~$\Tau(\bX_\Iend)\leq \ubar{c}(\text{S}(\bX_\Iend))$} 
		where, for~$\bar{\alpha},\,\ubar{\alpha}>0$ such \mbox{that $\bar{\alpha}+\ubar{\alpha}=\alpha$}, we have for all~$s'\in\mathcal{I}$
		\begin{align}
			\bar{c}(s')&=\min\left\{c\in \bar{\Tau}(\mathcal{X}_\text{S}(s')):\binom{\Iend}{s'}^{-1}\sum_{\bx_{\Iend}\in\mathcal{X}_\text{S}(s'): \;\Tau(\bx_{\Iend})\geq c} g_{{\Iend}}^\pi(\bx_{\Iend})\leq \bar{\alpha}\right\},\label{critval_cond_upper}
		\end{align}
		and~$\ubar{c}$ is defined similarly using the left tail and~$\ubar{\alpha}$~\citep[see, e.g.,][]{baas2025exact}. 
		In the above,~$\Tau(E)$ denotes the image of~$E\subseteq\mathcal{X}_{\Iend}$ under~$\Tau$, while~$\bar{\Tau}(E)=$\hbox{$\Tau(E)\cup\{-\infty,\infty\}$.}
	\end{definition}
	
	The next result, proven in~\citet{baas2025exact}, states that the conditional test based on~$\text{S}$ is exact under the model of the previous section, and will hence be denoted the \text{CX-$\text{S}$} test in the following. As the critical value of the CX-S test is based on the range of the test statistic, this result holds without restrictions on the test statistic function (although the choice of statistic does influence OCs such as power).

	\begin{lemma}\label{cor:cond_exact}
		For every parameter vector~$\btheta$ satisfying the null hypothesis~$H_0$ we have~$$\mathbb{P}^\pi_{\btheta}\Big(\Tau(\bX_{\Iend})\geq \bar{c}(\text{S}(\bX_{\Iend}))\text{ or  }\Tau(\bX_{\Iend})\leq \ubar{c}(\text{S}(\bX_{\Iend}))\Big)\leq \alpha.$$
	\end{lemma}
	
	\subsubsection{Unconditional exact test}
	
	In this subsection we discuss an unconditional test for RA designs, generalizing Barnard's test~\citep{barnard1945new}. The unconditional test uses a critical value that bounds the highest rejection rate under the null hypothesis by the significance level. \\
	
	\begin{definition}[(Unconditional test)]~\label{def:uncond_test}
		An unconditional test for  null hypothesis~$H_0$ given a test statistic function~$\Tau$,
		RA~procedure~$\pi$, and significance level~$0<\alpha<1$ 
		rejects when~$\Tau(\bX_{\Iend})\geq \bar{c}$ or~$\Tau(\bX_{\Iend})\leq\ubar{c}$ where, for~$0<\bar{\alpha},\,\ubar{\alpha}<1$ such 
		that~$\bar{\alpha}+\ubar{\alpha}=\alpha$, we have~(for~$\mathbb{P}^\pi_{\btheta}(\bX_{\Iend}=\bx_{\Iend})$ given in~\eqref{eqn:likelihood})
		\begin{align}
			\bar{c}&=\min\left\{c\in \bar{\Tau}(\mathcal{X}_{\Iend}):\max_{\substack{\btheta\in[0,1]^2,\\\theta_\C=\theta_\D}}\sum_{\bx_{\Iend}\in\mathcal{X}_{\Iend}:\,\Tau(\bx_{\Iend})\geq c}\mathbb{P}^\pi_{\btheta}(\bX_{\Iend}=\bx_{\Iend})\leq \bar{\alpha}\right\},
		\end{align}
		and~$\ubar{c}$ is defined similarly using the left tail and~$\ubar{\alpha}$~\citep[see, e.g.,][]{baas2025exact}.
	\end{definition}
	
	The next result, which follows immediately from Definition~\ref{def:uncond_test} as the maximum rejection rate over the null set bounds the rejection rate at any point in the null set, states that the unconditional test is exact under the model of Section~\ref{sect:model}. The unconditional test will be denoted by the \emph{UX test} in the following. 
	
	\begin{lemma}\label{cor:uncond_exact}
		Under~$H_0$ it holds \mbox{that 
			~$\mathbb{P}^\pi_{\btheta}\Big(\Tau(\bX_{\Iend})\geq\bar{c}\text{ or }\Tau(\bX_{\Iend})\leq \ubar{c}\Big)\leq \alpha,$}
		where~$\bar{c},\,\ubar{c}$ are as given in Definition~\ref{def:uncond_test}.
	\end{lemma}
	
	Algorithm~2 in~\citet{baas2025exact} can be used to calculate~$\bar{c},\ubar{c}$ up to a desired precision.
	
	The UX test is defined similarly to the commonly-used~\emph{calibrated} test, where the distribution of the test statistic under a parameter configuration~$ {\btheta'}$ under~$H_0$
	is used to determine a critical value. \\
	\begin{definition}[(Calibrated test)]~\label{def:calibrated_test}
		A calibrated test given a test statistic function~$\Tau$ and parameter vector~$ {\btheta'}\in[0,1]^2$ such that~$\theta'_\C=\theta'_\D$,
		RA~procedure~$\pi$, and significance level~$0<\alpha<1$ 
		rejects when~$\Tau(\bX_{\Iend})\geq \bar{c}$ or~$\Tau(\bX_{\Iend})\leq\ubar{c}$ where, for~$0<\bar{\alpha},\,\ubar{\alpha}<1$ such 
		that~$\bar{\alpha}+\ubar{\alpha}=\alpha$, we have
		\begin{align}
			\bar{c}&=\min\left\{c\in \bar{\Tau}(\mathcal{X}_{\Iend}):\sum_{\bx_{\Iend}\in\mathcal{X}_{\Iend}:\,\Tau(\bx_{\Iend})\geq c}\mathbb{P}^\pi_ {\btheta'}(\bX_{\Iend}=\bx_{\Iend})\leq \bar{\alpha}\right\},
		\end{align}
		and~$\ubar{c}$ is defined similarly using the left tail and~$\ubar{\alpha}$, while~$\mathbb{P}^\pi_ {\btheta'}(\bX_{\Iend}=\bx_{\Iend})$ is given in~\eqref{eqn:likelihood}.
	\end{definition}
	The calibrated test is often applied to ensure type I error rate control for the PPCS test under an assumed parameter vector~$ {\btheta'}$~\citep[see, e.g.,][]{du2018comparing, yannopoulos2020advanced,viele2020comparison2, viele2020comparison}.  As there is no type I error rate guarantee for this test when the true parameter vector is different from the parameter~$ {\btheta'}$ the test is calibrated for, the calibrated test is not exact under~$H_0$, though it can be seen as an exact test for the null hypothesis~$H_0':\btheta = {\btheta'}$. Although not considered in this paper, one can also calibrate a test to a strict subset~$\tilde{\Theta}_0\subset[0,1]$ which leads to an~(intermediate) exact test for all success rates in~$\tilde{\Theta}_0.$
	
	\subsection{Operating characteristics and their exact calculation}\label{OCscalc}
	Apart from exact tests, the likelihood~\eqref{eqn:likelihood} also allows for the calculation of exact operating characteristics~(OCs), such as the power or type I error rate.
	An OC can be written as~$\mathbb{E}_{\btheta}^\pi[f(\bX_{\Iend},\btheta)]$ for a function~$f:\mathcal{X}_{\Iend}\times [0,1]^2\mapsto\mathbb{R}$.
	The OCs considered in the paper will be:
	\begin{itemize}
		\item {\bf Rejection rate}:\\ This OC is calculated as $\mathbb{E}_{\btheta}^\pi[f(\bX_{\Iend},\btheta)]$ for~$f(\bx_{\Iend},\btheta)=\mathbb{I}\Big(\Tau(\bx_{\Iend})\leq \ubar{c}(\bx_{\Iend})\text{ or }\Tau(\bx_{\Iend})\geq \bar{c}(\bx_{\Iend})\Big)$. 
		For~$\delta\in(0,1)$ let~$\Theta_{\delta}=\{\btheta:\theta_\D-\theta_\C=\delta\}$, then this OC is the type I error rate when~$\btheta\in\Theta_0$ and power when~$\btheta\in\Theta_\delta$ for~$\delta\neq 0$ under the test based on test statistic~$\Tau$ and lower and upper critical value functions~$\ubar{c},\bar{c}$. For a significance level~$\alpha$ it is desired to have the type I error rate bounded by~$\alpha$, while higher power is better. We denote one minus the type I error rate by the true negative rate~(TNR).\\
		\item {\bf Expected
			proportion of allocations on the superior arm (EPASA)}:\\ This OC equals~$$
		\sum_a\mathbb{E}^\pi_\btheta[n_a(\bX_{\Iend})/\Iend]\mathbb{I}(\theta_a=\max_{a'}\theta_{a'})-\mathbb{I}(\theta_\C=\theta_\D)/2.$$ The OC represents the proportion of participants on the superior arm. Higher values of EPASA are better.\\
		\item {\bf Probability of an imbalance in the wrong direction~(PIWD($\varphi$))}:\\ For~$\btheta\in\Theta_\delta$ and~$\delta\neq 0$ this OC, also considered in~\citet{thall2015statistical}, equals~$$
		\mathbb{P}^\pi_\btheta(n_\C(\bX_{\Iend})/\Iend>n_\D(\bX_{\Iend})/\Iend+\varphi)\mathbb{I}(\theta_\D>\theta_\C)+\mathbb{P}^\pi_\btheta(n_\D(\bX_{\Iend})/\Iend>n_\C(\bX_{\Iend})/\Iend+\varphi)\mathbb{I}(\theta_\C>\theta_\D)$$ and represents the probability of allocating a proportion~$\varphi\in[0, 1]$ more participants to the inferior arm than to the superior arm. Lower values of this OC are better. We denote one minus PIWD($\varphi$) by the probability of no imbalance in the wrong direction~(PNIWD($\varphi$)).\\
		\item {\bf Bias of the treatment effect estimator}:\\ This OC equals~$$
		\mathbb{E}^\pi_\btheta\left[\hat{\theta}_\D(\bX_{\Iend})-\hat{\theta}_\C(\bX_{\Iend})\right]-({\theta}_\D-{\theta}_\C),$$
		where~$\hat{\theta}_a(\bX_{\Iend})= (s_{a}(\bX_{\Iend}) + \iota(\bX_{\Iend}))/(N_{a} + 2\iota(\bX_{\Iend})))$ and $\iota(\bX_{\Iend})=\mathbb{I}(\min_an_a(\bX_{\Iend})=0)$.  This OC represents the expected error in the treatment effect estimate~$\hat{\theta}_\D(\bX_{\Iend})-\hat{\theta}_\C(\bX_{\Iend}).$
	\end{itemize}

	We have from~\eqref{eqn:likelihood}:
	\begin{equation}\mathbb{E}_{\btheta}^\pi[f(\bX_{\Iend},\btheta)]=\sum_{\bx_{\Iend}\in\mathcal{X}_{\Iend}}f(\bx_{\Iend},\btheta)g^\pi_{\Iend}(\bx_{\Iend})\prod_{a\in\{\C,\D\}}\theta_a^{s_a(\bx_{\Iend})}(1-\theta_a)^{n_{a}(\bx_{\Iend})-s_a(\bx_{\Iend})},\label{eqn:OCcalc}\end{equation}
	which can be written as~$(\bg_{\Iend}^ \pi)^\top(\bff_{\btheta}\circ\bp_{\btheta})$ with~$\circ$ the Hadamard product and~$\bp_{\btheta}$ containing the product-term on the right in the above expression. Hence, we can store~$\bg_{\Iend}^\pi$ and then only have to take the inner product with~$(\bff_{\btheta}\circ\bp_{\btheta})$ for different vectors~$\btheta$ when we want to calculate~\eqref{eqn:OCcalc} for different values of~$\btheta$.
	
	While in practical applications one might have a more specific idea of the realistic parameter range, we aim to offer a robust picture to highlight the potential variation between parameters even with the same treatment effect difference $\delta$. Hence in the following, we discuss two measures that describe the behaviour of the OCs over the complete parameter space, namely the average over OCs and minimum/maximum over OCs.
	
	\subsubsection{Average over operating characteristics}
	Let~$|\Theta|$ be the area or length of~$\Theta\subset[0,1]^2$. 
	Based on~\eqref{eqn:OCcalc} the average of the OC~$\mathbb{E}_{\btheta}^\pi[f(\bX_\Iend,\btheta)]$ over~$\Theta$ is represented by~$\mathbb{E}_{\Theta}^\pi[f(\bX_\Iend,\btheta)]$ and defined as
	\begin{align}&\frac{1}{|\Theta|}\int_{\Theta}\mathbb{E}_{\btheta}^\pi[f(\bX_{\Iend},\btheta)]d\btheta\nonumber = \sum_{\bx_{\Iend}\in\mathcal{X}_{\Iend}}f(\bx_{\Iend},\btheta)g^\pi_{\Iend}(\bx_{\Iend})\underbrace{\frac{1}{|\Theta|}\int_{\Theta}\prod_{a\in\{\C,\D\}}\theta_a^{s_a(\bx_{\Iend})}(1-\theta_a)^{n_{a}(\bx_{\Iend})-s_a(\bx_{\Iend})}d\btheta}_{\bp_{\Theta}(\bx_\Iend)}\nonumber\\&= \sum_{\bx_{\Iend}\in\mathcal{X}_{\Iend}}f(\bx_{\Iend},\btheta)g^\pi_{\Iend}(\bx_{\Iend})\bp_{\Theta}(\bx_\Iend).\label{eqn:avg_OC}\end{align}
	For instance,  we have
	\begin{align}
		\bp_{\Theta_{\delta}}(\bx_\Iend)&=\int_{0}^{1-\delta}\frac{\prod_{a\in\{\C,\D\}}(\theta_\C+\delta\cdot\mathbb{I}(a=\D))^{s_a(\bx_{\Iend})}(1-\theta_\C-\delta\cdot\mathbb{I}(a=\D))^{n_a(\bx_{\Iend})-s_a(\bx_{\Iend})}}{1-\delta}d\theta_\C
		\label{eqn:OCdiff},\\
		\bp_{\Theta_{0}}(\bx_\Iend)&=\text{B}(s(\bx_{\Iend})+1,\Iend-s(\bx_{
			\Iend
		})+1),\label{eqn:OCHOavg}
	\end{align}\\
	where~B is the Beta function.
	The average OC generalizes the OC at a single parameter vector~(as we can \hbox{take~$\Theta=\{\btheta\}$)} and represents the average value of the OC over a specific part of the parameter space.  It hence better represents the behaviour of the OC under RA procedure on average over the part of the parameter space that is of interest. Although not interpreted as a Bayesian measure, the average OC equals the Bayesian average value~$\int_{[0,1]^2}\mathbb{E}_{\btheta}^\pi[f(\bX_{\Iend},\btheta)]p(\btheta)d\btheta$ of the OC for a prior density~$p$ on~$[0,1]^2$ where~$p$ equals the uniform prior on~$\Theta_\delta$~(i.e.,~$p(\btheta)=\mathbb{I}(\theta\in\Theta_\delta)/|\Theta_\delta|$). Using the average power or type~I error rate to objectively evaluate the performance of a statistical test in the case where there is no prior information, or as a long-term evaluation measure, has been argued for in, e.g.,~\citet{rice1988new},~\citet{ANDRES1994555}, and~\citet{Best31052024} where the first and last paper propose the average OC in a more general Bayesian context.
	\subsubsection{Grid-approximated minimum and maximum over operating characteristics}
	In order to find the grid-approximated minimum and maximum OCs, we discretize the set~$\Theta_\delta$ to a finite set~$\hat{\Theta}_\delta$.
	The minimum OC~$\mathbb{E}_{\btheta}^\pi[f(\bX_{\bar{i}},\btheta)]$ over~$\hat{\Theta}_\delta$~(approximating the minimum OC over~$\Theta_\delta$) equals~$\min_{\btheta\in\hat{\Theta}_{\delta}}\mathbb{E}_{\btheta}^{\pi}[f(\bX_{\bar{i}},\btheta)]$. The maximum OC is calculated in a similar vein. The minimum power to indicate the worst-case behaviour of a test has also been considered  in~\citet{Haber01011987} although not in addition to the average.
	\section{The effect of the burn-in length in a Bayesian response-adaptive design} \label{sect:results}
	
	In this section, we investigate the effect of the burn-in length on the type I error rate  for the calibrated test based on the PPCS~(Section~\ref{typeIerrorcurves}),  we consider what the added value of using a burn-in period is when using an exact test in a BRAR design~(Section~\ref{sect:exact_tests}), consider how EPASA and PIWD(0.1) change with the burn-in length~(Section~\ref{sect:participant}),  evaluate the effect of the burn-in length on treatment effect estimation bias~(Section~\ref{sect:bias_results}),
	give a recommendation for choosing the burn-in length~(Section~\ref{sect:recommend}), and provide a sensitivity analysis with respect to the prior used for BRAR, where our default is the uniform prior~(Section~\ref{sect:sensitivity_analysis}).
	As indicated in Section~\ref{sect:intro}, to purely investigate the effect of the burn-in period length, we limit the focus on two-arm BRAR designs using a burn-in period as the sole tuning parameter, while the trial application in Section~\ref{sect:ARREST} considers a trial with blocked allocation, early stopping, and clipped allocation probabilities. 
	
	We consider two specific trial sizes $\bar{i}\in\{60,240\}$, which could represent an early-stage exploratory trial and a small confirmatory trial, respectively. These two numbers were chosen due to their large number of divisors, making them more suitable and probable to be used in designs using blocked allocation than, e.g., trial sizes~50 and 250. Following~\citet{thall2015statistical}, we take~$\varphi=0.1$ to define PIWD.
	In the remainder, we {set~$\hat{\Theta}_\delta=\{\btheta\in[0,1]^2:\theta_\C=\theta_\D-\delta,\;\theta_\D\in\{\delta,\delta+0.01,\dots,1.00\}\}$} and~$\ubar{\alpha}=\bar{\alpha}=0.025$.  All calculations in this paper are exact and not simulation-based. 
	
	The allocation probabilities~$\pi^{\text{B}}_b$ were calculated using Gauss–Kronrod quadrature using the \verb|QuadGK| \verb|Julia| package~\citep[see the][]{quadgkpackage} with absolute tolerance~$10^{-3}$. Computation of the allocation probabilities for all states~$\bx\in\mathcal{X}$ for~$\Iend=240$ took~2,950 seconds
	on a standard laptop~(1.7 Ghz, 10 cores, 32 GB RAM). This vector can also be used for~$\Iend=60$ and different burn-in parameters~$b$, hence this vector only needs to be calculated once.  
	Computation of~$g_{\bar{i}}^{\pi_b^{B}}$, where we loop over the same set of states but perform a simpler calculation than numerical integration, took~$173$ seconds for~$\Iend=240$ and~$b=0$~(which is the value of~$b$ with the longest computation time).
	The Gauss–Kronrod quadrature with a~(default) relative tolerance $\sqrt{\epsilon}$~(where $\epsilon$ is the machine epsilon in Julia, e.g., for one device this was around $2.2\cdot10^{-16}$) was used to compute the values~$p_{\bTheta_{\delta}}(\bx_{\Iend})$ for every state~$\bx_{\Iend}\in\mathcal{X}_{\Iend}$ \hbox{and~$\delta\in\{0.1,0.2,0.4\}$, }while~\eqref{eqn:OCHOavg} was used to compute average OCs under~$H_0$. 
	For~$\bar{i}=240$ the former calculations took~46
	, 44
	, and 40
	seconds, respectively, while the latter calculation took~2 
	seconds~(as no numerical integration is needed).
	The PPCS statistic for each final state was calculated with an absolute tolerance~$10^{-6}$, which took 183 
	seconds for~$\Iend=240$.
	Based on~\citet{jacko2019binarybandit} the amount of values~$\pi_b^B, g^{\pi_b^B}_{i}$ to calculate~(equal to~$|\mathcal{X}|$) is of order~$\mathcal{O}(\Iend^4)$, while the amount of end-states~(hence the amount of average probabilities and PPS values to calculate) grows with order~$\mathcal{O}(\Iend^3)$.

	\subsection{Issues arising from the application of commonly used tests for BRAR designs} \label{typeIerrorcurves}
	We will consider the calibrated test based on the PPCS defined in Section~\ref{sect:exact}, where we calibrate this test to the parameter configuration~$\theta_\C=\theta_\D=0.5$ which induces the highest outcome variance under~$H_0$. 
	
\begin{figure}[h!]
		\centering
		\includegraphics[width=\linewidth]{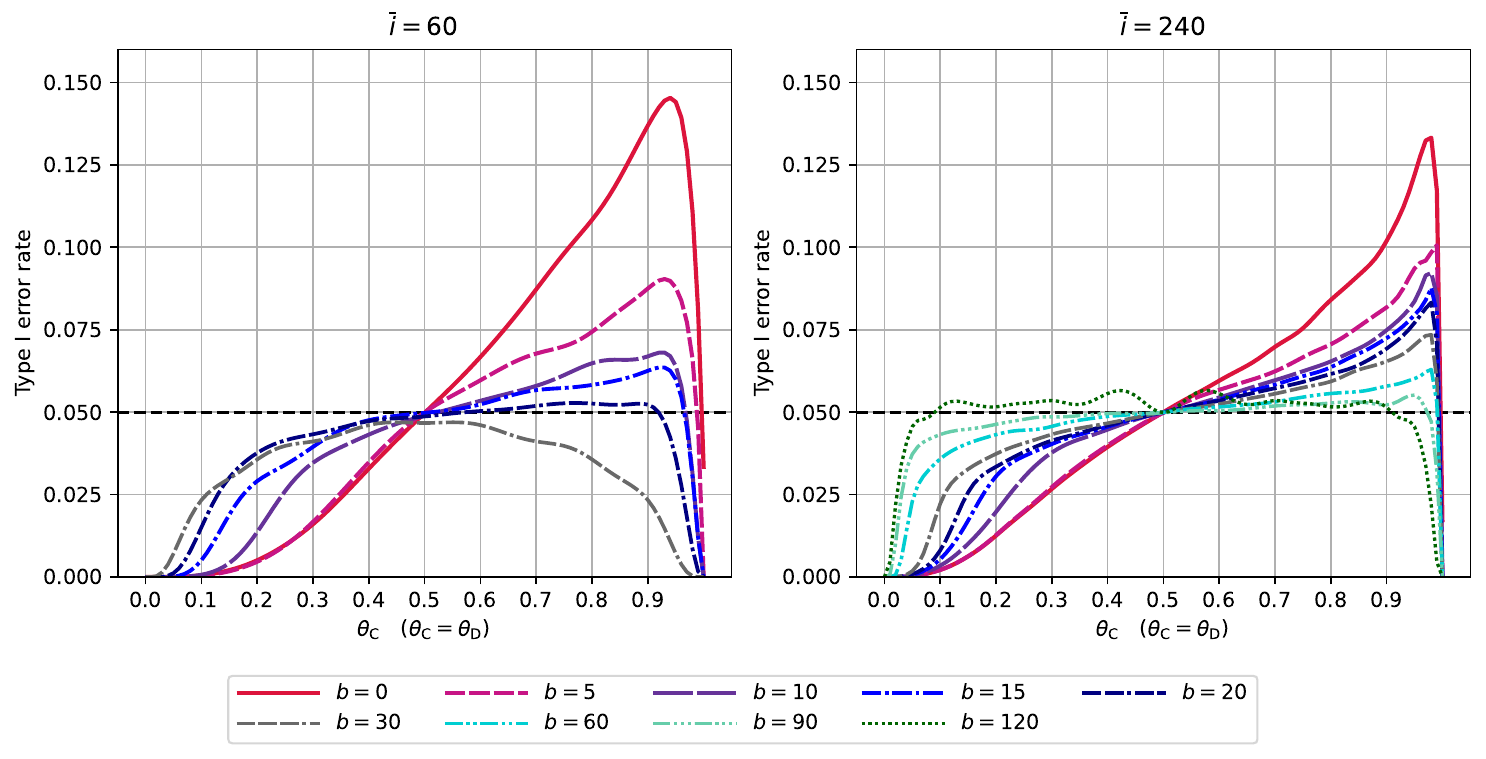}
		\caption{Type I error rate profiles for the Bayesian response-adaptive randomization design with calibrated test (calibrated for~$\theta_\C=\theta_\D=0.5$) based on the posterior probability of control superiority for trial sizes~$\bar{i}=60$ and~$\bar{i}=240$, across~$\theta_\C=\theta_\D\in \{0.00,0.01,\dots,1.00\}$ for different burn-in lengths~$b$. The significance level~$\alpha$ was set to~$0.05$ (indicated by the horizontal dashed line).}
		\label{fig:t1_profile_PPS}
	\end{figure}

	Figure~\ref{fig:t1_profile_PPS} shows the type I error rate profile for the calibrated test for RA procedure~$\pi^\text{B}_b$. When no burn-in is used, the calibrated critical value does not control the type I error rate well, reaching a type I error rate around 14.53\% for common success rates above~$0.9$ and~$\Iend=60$, almost three times the nominal significance level.  Designs with a larger burn-in, e.g., those where~$b\geq \Iend/4$, show a more balanced type~I error rate profile reaching a lower maximum value. For~$\Iend=60$ the type I error rate is under control when~$b=\Iend/2$ for all evaluated parameter values under the null, while this is not the case for~$\Iend = 240$.

	\FloatBarrier
	\FloatBarrier
	
	The variability of the type I error rates over different common success rates can be explained by the discreteness of the binary PPCS test coupled with the relatively small trial sizes considered. The asymmetry present for low burn-in lengths~$b$ can be explained intuitively. The PPCS will be close to 1/2 when the common success rate is small since the RA procedure is likely to switch between treatments when a failure is recorded~(inducing balance). On the other hand, for high common success rates, we are more likely to~(erroneously) favour one treatment because we keep on recording successes for that treatment. In cases where almost all successes are on one arm, the PPCS roughly equals the common success rate or one minus this value~(one of the success rates has a uniform distribution, while the other distribution has low variance) hence the rejection rate grows in the common success rate. 
	
	To keep the type I error rate roughly under 6\%, Figure~\ref{fig:t1_profile_PPS} suggests the burn-in length should be more than a quarter of the trial size~(i.e.,~$b> \Iend/4$). 
	For~$\Iend\geq60$ we show that this indeed seems to be a valid rule of thumb in the case of a two-arm BRAR and when testing~$H_0$ using the calibrated PPCS test. Table~\ref{tab:maxAStype1_PPS} shows the grid-approximated maximum type I error rate of the calibrated PPCS test versus~$b$ for~$\bar{i}\in\{20,40,60,80,100,240\}$. Table~\ref{tab:maxAStype1_PPS} shows that when the burn-in proportion~(BP) is higher than or equal to~$0.6$,  the type I error rate is controlled at~$6\%$ for~$\Iend\geq60$.

	\begin{table}[h!]
		\caption{Maximum type I error rate~(in \%) across the whole parameter space~(using a grid approximation)  of the calibrated test (calibrated for~$\theta_\C=\theta_\D=0.5$) based on the posterior probability of control superiority~(significance level~$\alpha=0.05$) under the fully sequential BRAR design as the burn-in length varies. Burn-in proportion~(BP) is the proportion of the trial within the burn-in stage, given by BP$=2b/\bar{i}$. BP$=1.0$ corresponds to equal allocation. Type I error rates above~$6\% $ are indicated in bold, with the severity of the inflation emphasized by a gradient~(red being worst).}\label{tab:maxAStype1_PPS}
		\centering
		\small
		\begin{tabular}{lccccccccccc}
\toprule
BP & 0 & 0.10 & 0.20 & 0.30 & 0.40 & 0.50 & 0.60 & 0.70 & 0.80 & 0.90 & 1.00 \\
\midrule
$\Iend=$ 20 & \textcolor{red}{\textbf{12.72}} & \textcolor{red}{\textbf{13.71}} & \textcolor{red}{\textbf{10.68}} & \textcolor{orange}{\textbf{9.54}} & \textcolor{orange}{\textbf{8.02}} & 5.69 & \textcolor{brown}{\textbf{6.20}} & 5.13 & \textcolor{brown}{\textbf{6.22}} & 5.43 & 5.00 \\
$\Iend=$ 40 & \textcolor{red}{\textbf{14.59}} & \textcolor{red}{\textbf{12.48}} & \textcolor{orange}{\textbf{8.55}} & \textcolor{brown}{\textbf{7.07}} & \textcolor{brown}{\textbf{6.23}} & \textcolor{brown}{\textbf{6.39}} & \textcolor{brown}{\textbf{6.02}} & 5.84 & 5.40 & 4.91 & \textcolor{brown}{\textbf{6.09}} \\
$\Iend=$ 60 & \textcolor{red}{\textbf{14.53}} & \textcolor{red}{\textbf{11.04}} & \textcolor{orange}{\textbf{8.59}} & \textcolor{brown}{\textbf{7.20}} & \textcolor{brown}{\textbf{6.76}} & \textcolor{brown}{\textbf{6.36}} & 5.47 & 5.62 & 5.15 & 4.99 & 4.69 \\
$\Iend=$ 80 & \textcolor{red}{\textbf{15.07}} & \textcolor{orange}{\textbf{9.83}} & \textcolor{orange}{\textbf{8.37}} & \textcolor{brown}{\textbf{7.42}} & \textcolor{brown}{\textbf{6.60}} & \textcolor{brown}{\textbf{6.21}} & 5.72 & 5.31 & 5.47 & 5.23 & 5.04 \\
$\Iend=$ 100 & \textcolor{red}{\textbf{14.56}} & \textcolor{orange}{\textbf{9.44}} & \textcolor{orange}{\textbf{8.10}} & \textcolor{brown}{\textbf{7.23}} & \textcolor{brown}{\textbf{6.95}} & \textcolor{brown}{\textbf{6.29}} & 5.78 & 5.49 & 5.62 & 5.23 & 5.07 \\
$\Iend=$ 240 & \textcolor{red}{\textbf{13.31}} & \textcolor{orange}{\textbf{9.01}} & \textcolor{brown}{\textbf{7.82}} & \textcolor{brown}{\textbf{7.23}} & \textcolor{brown}{\textbf{6.64}} & \textcolor{brown}{\textbf{6.30}} & 5.64 & 5.38 & 5.17 & 5.09 & 5.66 \\
\bottomrule
		\end{tabular}
	\end{table}
	\FloatBarrier
	In conclusion, the use of calibrated tests leads to a substantial risk of type I error inflation under parameter misspecification, and this problem cannot be fully eliminated by increasing the burn-in length. Due to regulatory demands for strict type I error rate control, 
	we study burn-in length's effect on OCs with an exact test.
	
	Appendix \ref{RR_wald} shows the same evaluation when instead of considering the PPCS as the test statistic, we use the Wald statistic with the Agresti-Caffo adjustment as defined in~\citet[Equation~(14)]{baas2025exact}. 
	We use a standard asymptotic two-sided Wald test~(significance level~5\%) instead of a calibrated test, common in theory but rare in BRAR trials with burn-in.~(see, e.g., Table~\ref{tab:maxAStype1}).  The finite-sample and asymptotic properties of the Wald test under RA procedures are discussed in~\citet{baldi2022simple}.

	The main differences between the results for PPCS and the Wald statistic are that the type I error rate profile for the Wald test is more symmetrical, while the maximum type I error rate inflation for~$b=0$ is slightly lower than for PPCS, around 10\%-12\%~(Table~\ref{fig:AS_N=60,240}), due to the more symmetrical type I error rate profile. 
	The commonalities with the PPCS test are that the rule~$b>\Iend/4$ works for controlling the type I error rate at~$6\%$ (in this case, even for all values of~$\Iend$, see Table~\ref{tab:maxAStype1}).

	\FloatBarrier
	\subsection{The added value of a burn-in when exact tests are applied in BRAR designs} \label{sect:exact_tests}
	This section presents the results for the exact tests given in Section~\ref{sect:exact} using the  PPCS statistic under RA procedure~$\pi^\text{B}_b$ for different burn-in lengths~$b$.
	Note that for each burn-in length~$b$ and type of exact 
	test, a different critical value is derived. To give some intuition, we provide some of these critical values in Appendix~\ref{app:critvals}, where we also include critical values for the Wald test.
	We provide a power comparison of the CX-S test and UX test, as both of them control the type I error rate.
	
\begin{figure}[h!]
		\centering
		\includegraphics[width=.99\linewidth]{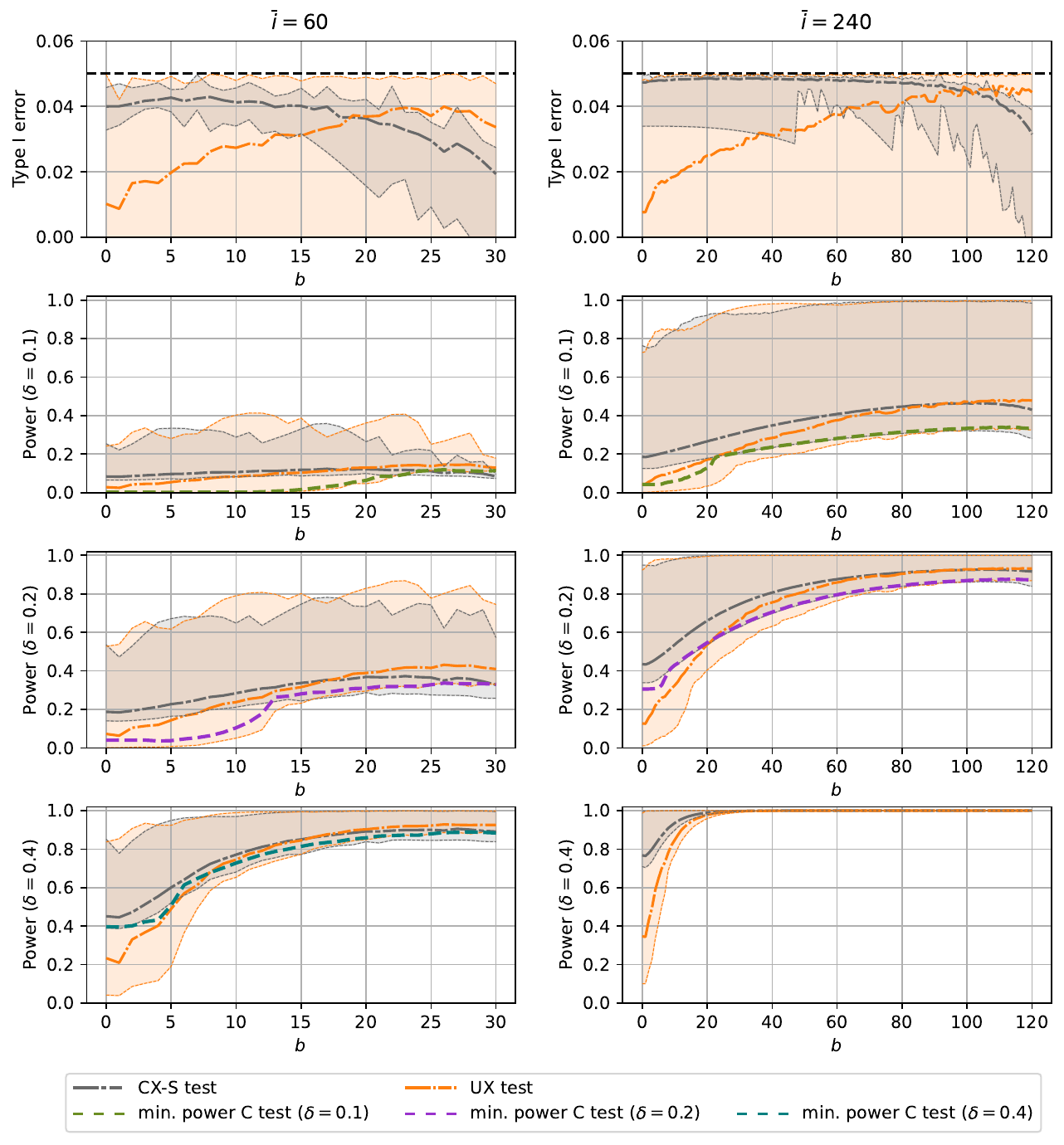}
		\caption{Type I error rate and power under treatment effects~$\delta\in \{0.1,0.2,0.4\}$ for   the Bayesian response-adaptive randomization design using the unconditional exact~(UX) test and conditional exact test based on total successes~(CX-S test) based on the posterior probability of control superiority with trial size~$\Iend=60$ and~$\Iend=240$ across different burn-in lengths. For comparison, we have plotted the~(grid-approximated) minimum power for the calibrated~(C) test for~$\Iend=60$, $\delta\in\{0.1,0.2,0.4\}$ and~$\Iend=240$, $\delta\in\{0.1,0.2\}$.
			The averages for each treatment effect~$\delta$ and fixed burn-in length are represented by thick lines, while the minimum and maximum are presented by thin dotted lines and the ribbons.  The significance level~$\alpha$ was set to~$0.05$.}
		\label{fig:CXUX_PPS}
	\end{figure}

	The graphs in the first row of Figure~\ref{fig:CXUX_PPS} show the average, and grid-approximated minimum and maximum type I error rate across different burn-in lengths. 
	As expected with exact tests, both the UX and CX-S tests maintain strict type I error control at 5\% across all burn-in lengths, ensuring a maximum type I error rate of 5\%. As $b$ increases, the average type I error rate for the UX test roughly increases, while it roughly decreases for the CX-S test. Hence, the UX (CX-S) test is likely overly conservative when the burn-in  length is small (large).  The minimum and maximum type I error rate for the CX-S test shows a highly non-smooth behaviour, which follows from the~{fact that the critical values for these tests are different for each burn-in length, and by the} discreteness of {binary} test{s in general}. 
	Increasing~$b$ exacerbates {this behaviour}, leading to a narrower range of treatment group sizes. 
	Figure~\ref{fig:CXUX_PPS}  shows that the type I error rate for the UX test varies between 0\% and 5\% for all burn-in lengths, yet the CX-S test shows much less variability over the parameter space and is less conservative than the UX test for smaller burn-in lengths, where the minimum type I error rate is around or higher than 3\% for~$b$ up to~$\Iend/4$.
	The maximum type I error rate for the CX-S test decreases for larger values of~$b$, where the maximum type I error rate for CX-S at~$b=\Iend/2$ roughly equals the average type I error rate of the UX test for~$b=\Iend/4$ for both~$\Iend\in\{60,240\},$ hence the CX-S test is likely overly conservative for larger burn-in lengths.
	We note that for~$b=\Iend/2$, the CX-S test equals Fisher's exact test~\citep{Fisher1934method}, while the UX test equals Barnard's test~\citep{barnard1945new}. Fisher's exact test is known to be more conservative than Barnard's test in the case of equal treatment group sizes and relatively small sample sizes~\citep{mehrotra2003cautionary}.
	
	The three bottom rows in Figure~\ref{fig:CXUX_PPS} show how the minimum, maximum and average power to under treatment effects~$\delta\in\{0.1,0.2,0.4\}$ vary across different burn-in lengths.
	The average power for~$b=\Iend/2$ is higher than for~$b=0$ for both exact tests, however, the average power does not increase monotonically in the burn-in length. For the CX-S test, the average power can furthermore attain a maximum at~$b<\Iend/2$, e.g., when~$\Iend=240$ the maximum occurs around~$b=100.$
	
	Figure~\ref{fig:CXUX_PPS} shows that for~$\Iend=60$  the average power for the CX-S test is higher than that of the UX test when $b<\bar{i}/4$, while it is lower for
	$b\geq \bar{i}/4$. We note that, generally, the power differences between the two tests are smaller in the latter situation~($b\geq \bar{i}/4$) than in the former situation~($b< \bar{i}/4$). 
	For~$\Iend = 240$ and~$\delta\in\{0.1,0.2\}$ the CX-S test has higher
	average and minimum power up to~$b=80$.
	This agrees with~\citet{baas2025exact}, showing  CX-S test's  higher power over the UX test for many RA procedures with a high degree of response-adaptiveness. 
	A potential explanation of this phenomenon lies in the high dependency (for more aggressive RA procedures) of
	the distribution of treatment group sizes on 
	total successes.  
	On the contrary, for larger burn-in lengths this is not the case and the power of the CX-S test suffers from the higher discreteness of the conditional distribution of the test statistic. 
	Figure~\ref{fig:CXUX_PPS} shows a large spread in the power for the UX test for, e.g,~$\delta=0.2$ and~$\Iend=60$ and~$\delta=0.1$ and~$\Iend=240$, as indicated by the maximum and minimum power values. This 
	could be due to the highly asymmetrical type I error rate profile for the PPCS test, 
	hence 
	the UX
	PPCS
	test 
	is
	overly conservative for certain 
	parameter values. 
	Lastly, Figure~\ref{fig:CXUX_PPS} shows that
	the minimum power for the CX-S test is higher than the  minimum power for the calibrated test for at least a few burn-in lengths~(indicated  by dashed lines).
	\FloatBarrier
	
	\noindent This outperformance mainly happens for low success rates, where the calibration test was also shown to be conservative. 
	Hence, the CX-S test has, for some parameter configurations, higher power than the commonly used calibration test with the added benefit of being exact.

	Figure~\ref{fig:typeIall} shows the evaluation above for the Wald test instead of the PPCS test.  
	As the type I error rate profile of the asymptotic Wald test is more regular than that of the calibrated PPCS test for small burn-in lengths, the UX Wald test is less conservative for small burn-in lengths than the UX PPCS test. Comparing the CX-S Wald and CX-S PPCS tests, the type I error results are very similar.
	The \hbox{CX-S} Wald test often shows higher maximum power than the UX Wald test, while less often showing higher minimum power than the asymptotic Wald test~(only for~$\delta=0.10,\Iend=60$) than under the calibration PPCS test comparison. 
	The CX-S test again shows higher average power than the UX test when $b<\bar{i}/4$ and vice versa for $b\geq \bar{i}/4$, and maximum average power values are again found for~$b<\Iend/2.$
	
	\subsubsection{Optimal burn-in proportion across the parameter space} 
	
\begin{figure}[h!]
		\centering
		\includegraphics[width=.65\linewidth]{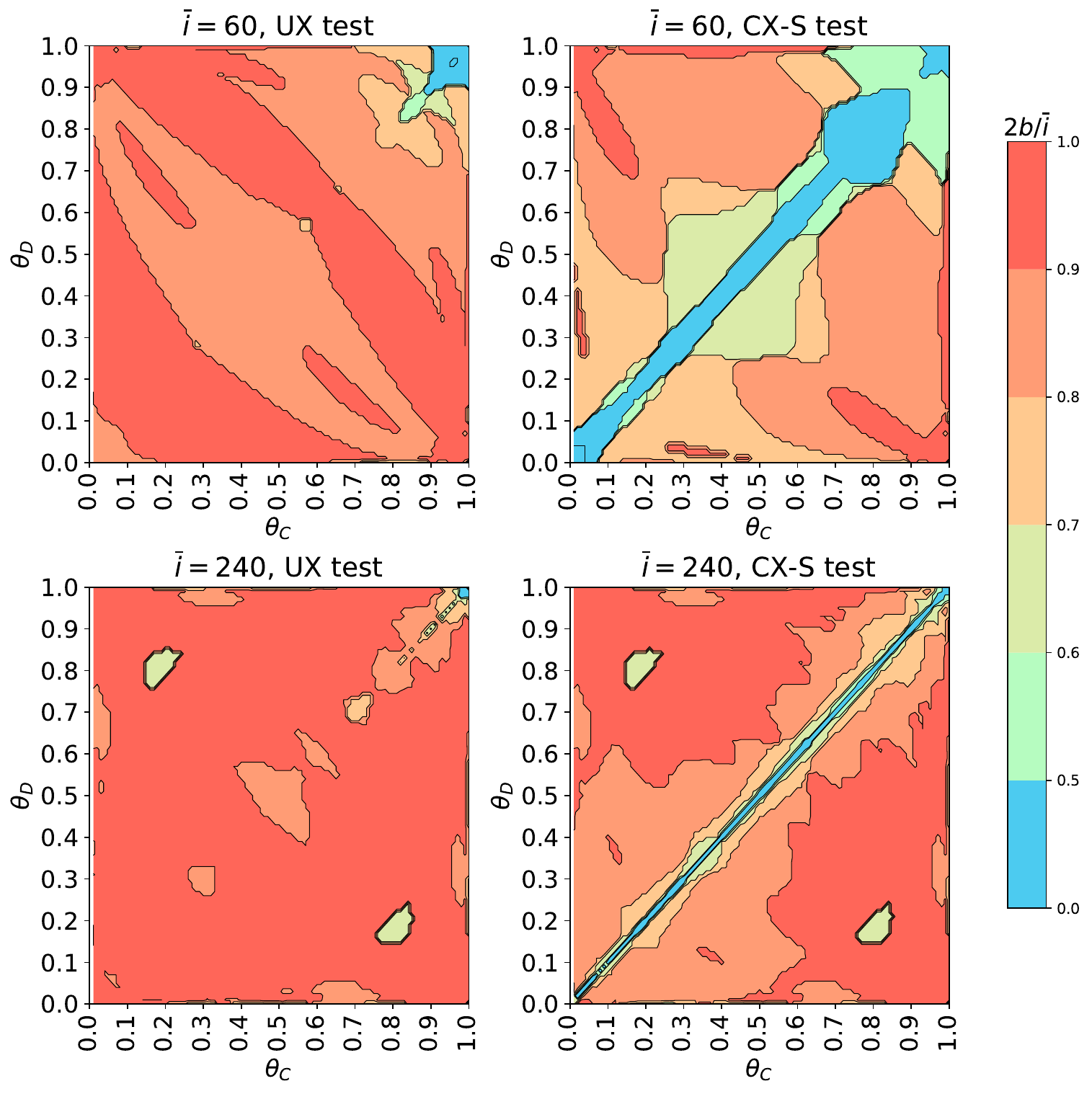}
		\caption{Optimal burn-in proportions~($2b/\Iend$) in terms of power of the Bayesian response-adaptive randomization design using the unconditional exact~(UX) and conditional exact test based on total successes~(CX-S test) based on the posterior probability of control superiority for every parameter configuration in the grid~$\theta_\C,\theta_\D\in\{0.00,0.01,\dots, 1.00\}.$ The considered trial sizes are~$\Iend=60$ and $\Iend=240$ participants.}
		\label{fig:optimal_burnin_PPCS}
	\end{figure}

	Figure~\ref{fig:optimal_burnin_PPCS} shows the \emph{optimal burn-in proportion  in terms of power}~(P-OBP) for the UX and CX-S test for each parameter \hbox{configuration~$\theta_\C,\theta_\D\in\{0.00,0.01,\dots,1.00\}$.} 
	
	\FloatBarrier
	\FloatBarrier
	
	\noindent As the calibrated test does not control type I error rates for every parameter configuration, this test was not considered in this evaluation. Note that the P-OBP can only be exactly computed using our calculation method, finding these maxima using simulation is prohibitive through Monte Carlo errors combined with the fluctuating and non-smooth behaviour of the power in the burn-in length.  
	
	The patterns in Figure~\ref{fig:optimal_burnin_PPCS} are hard to explain. Due to the discreteness of the binary tests, with changing critical values as the burn-in length changes, the power curve as a function of the burn-in length has a highly non-smooth behaviour and can suddenly jump to a high value.
	What first stands out in Figure~\ref{fig:optimal_burnin_PPCS} is that the P-OBP for power is often less than~$1.0$ for~$\Iend=60$. For the UX test it is often higher than~0.8, whereas for the CX-S test the P-OBP is often lower~(especially around the diagonal). For~$\Iend=240$ the P-OBPs are higher; the values for the UX test are closer to 1.0 than for~$\Iend=60$ and the values for the CX-S test are again lower~(especially around the diagonal). 
	
	Figure~\ref{fig:optimal_burnin_Wald} contains the P-OBP plots when the Wald statistic is used instead of PPCS, where overall the findings are the same as for PPCS. The subfigures for the CX-S test are very similar to the ones in Figure~\ref{fig:optimal_burnin_PPCS} where, upon inspecting the results, numerical differences were seen although they were very small. This could be explained by the fact that the CX-S test fixes total successes, hence the behaviour of this test is less sensitive to the choice of the test statistic. Larger differences are indeed seen for the UX test, where for~$\Iend=60$ lower P-OBPs are seen for the Wald test than for the PPCS test. 
	In conclusion, for the considered trial sizes the P-OBP for power depends heavily on the choice of test and is often not equal to one.
	
	\subsection{Impact of burn-in length on participant benefit metrics} \label{sect:participant}
	One of the the main arguments in the literature for using RA procedures is the potential to allocate more participants to the better treatment. This section investigates the behaviour of EPASA and PIWD(0.1) as we vary the burn-in length. Note these metrics are not inference-related, hence test-independent.
	
	\subsubsection{EPASA and PIWD(0.1) for different burn-in lengths}
	\begin{figure}[h!]
		\centering
		\includegraphics[width=\linewidth]{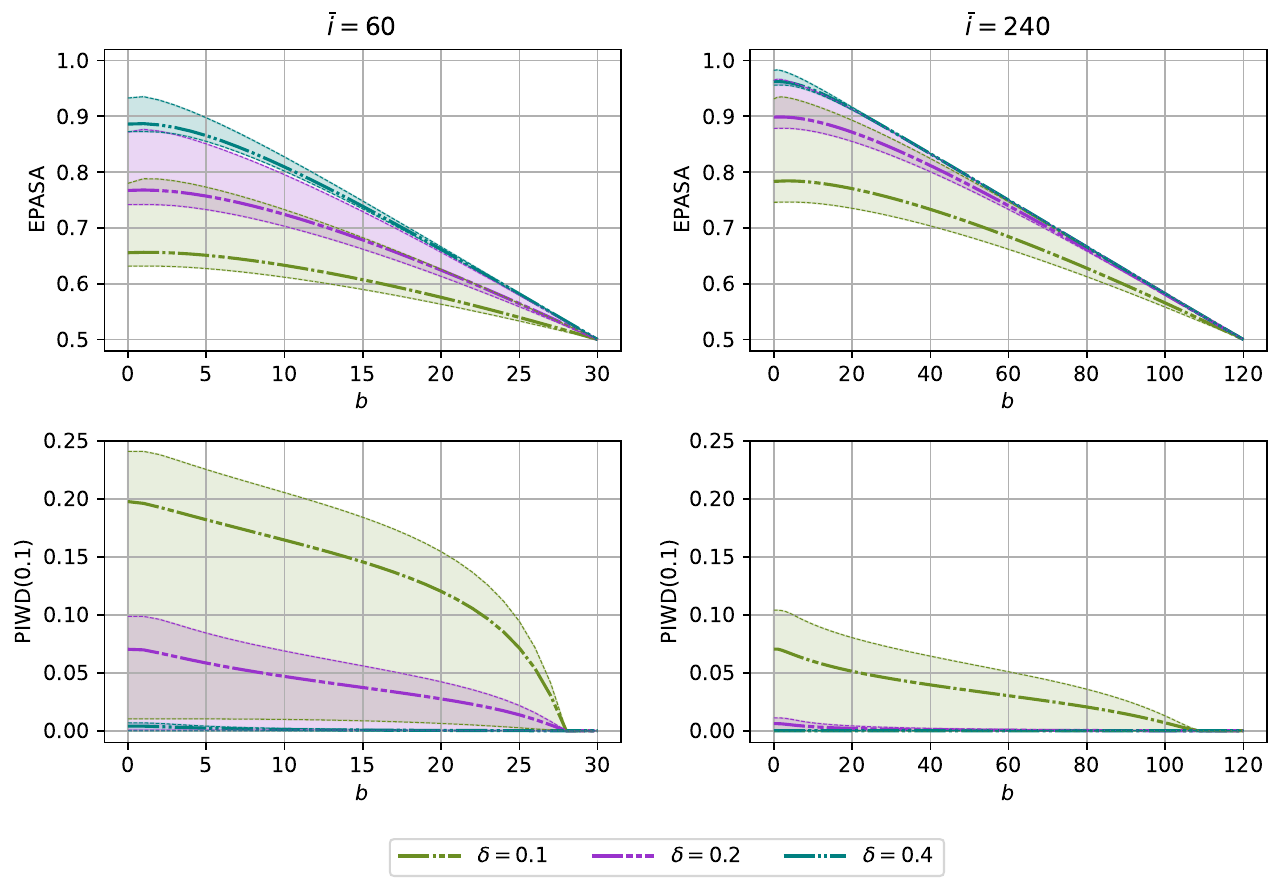}
		\caption{Expected proportion of allocations on the superior arm (EPASA) and probability of an imbalance in the wrong direction~(PIWD(0.1)) under different burn-in lengths~$b$ and treatment effects~$\delta\in\{0.1,0.2,0.4\}$.The averages for each treatment effect~$\delta$ and fixed burn-in length are represented by thick lines, while the minimum and maximum are presented by thin dotted lines and the ribbons.}
		\label{fig:imbl_epasa}
	\end{figure}
	
	The top row in Figure~\ref{fig:imbl_epasa} shows the average and grid-approximated minimum and maximum EPASA 
	versus
	the burn-in length for treatment \hbox{effects~$\delta\in\{0.1,0.2,0.4\}$.} 
	As expected, for every treatment effect~$\delta$, lower burn-in lengths give a higher average, minimum and maximum EPASA, and increasing the sample size and increasing~$\delta$ raises EPASA. A surprising observation here is that EPASA does not necessarily decrease as the burn-in length increases, e.g., the maximum average EPASA occurs for~$b=1$ when~$\bar{i}=60, \delta=0.1$.  The average, minimum and maximum EPASA increases initially, then decreases slowly, and then it decreases linearly as~$b$ increases. 
	
	One explanation is
	that a longer burn-in will more likely identify the better arm, leading to almost all participants being allocated to the best arm in the response-adaptive phase, tracing out a straight line on the EPASA graph.
	
	The bottom row in Figure~\ref{fig:imbl_epasa} shows that the average and grid-approximated minimum and maximum~PIWD(0.1) decrease as we increase the burn-in length. 
	The lines are in agreement with~\citet{Robertson_Lee_López-Kolkovska_Villar_2023}, who state that more aggressive RA procedures are more likely to have higher probabilities of imbalances in the wrong direction, which in our case would correspond to a lower burn-in length.
	As expected, for any fixed~$\delta$, lower burn-in lengths give a higher PIWD(0.1), while increasing sample size and increasing~$\delta$ reduces this OC. Initially for small burn-in lengths, PIWD(0.1) decreases in a roughly linear fashion, then towards the end, the PIWD(0.1) drops steeply towards 0. This is because for~$b> \Iend(1-\varphi)/2$ it is impossible to achieve imbalance in the wrong direction.
	
	For both metrics, the variation over the parameter space for~$\delta=0.1$ is larger than the variation \hbox{for~$\delta\in\{0.2,0.4\}$.} It is difficult for the RA procedure to detect a small treatment effect. That said, for~$\delta\geq 0.2$ and $\bar{i}=240$ the PIWD(0.1) is less than~$1\%$ regardless of the burn-in length, so we almost certainly improve in-trial participant benefit when the treatment effect is not too small given the trial size.

	While it may seem alarming that the average PIWD(0.1) for $\bar{i}=60, \delta=0.1$ can get as high as 20\% for small burn-in lengths, we highlight that the interpretation of the PIWD metric is not straightforward. 
	For instance, for $\Iend=60,\theta_{\C}=0.4,\theta_{\D}=0.5$ we find~PIWD$(0.1)=0.22$, meaning that the probability of having at least 33 participants allocated to the control treatment, leading to at least~$0.3$ 
	more expected treatment failures in total than under fixed equal allocation, is 22\%.
	The severity of this imbalance still depends on the probability of having an even worse misallocation, such as 65\% of participants being allocated to the worst arm. If this is zero, then the situation would not be severe after all, since for fixed equal allocation and~$\delta=0.1$  the probability of having 10\% less expected successes for 50\% of participants is 100\%. Hence, although we are only considering one imbalance measure, we recommend looking at different imbalance measures when choosing a burn-in length~(e.g., PIWD$(\varphi)$ for different values~$\varphi$).
	
	\subsubsection{PIWD(0.1) across the parameter space}
\begin{figure}[h!]
		\centering
		\includegraphics[width=0.65\linewidth]{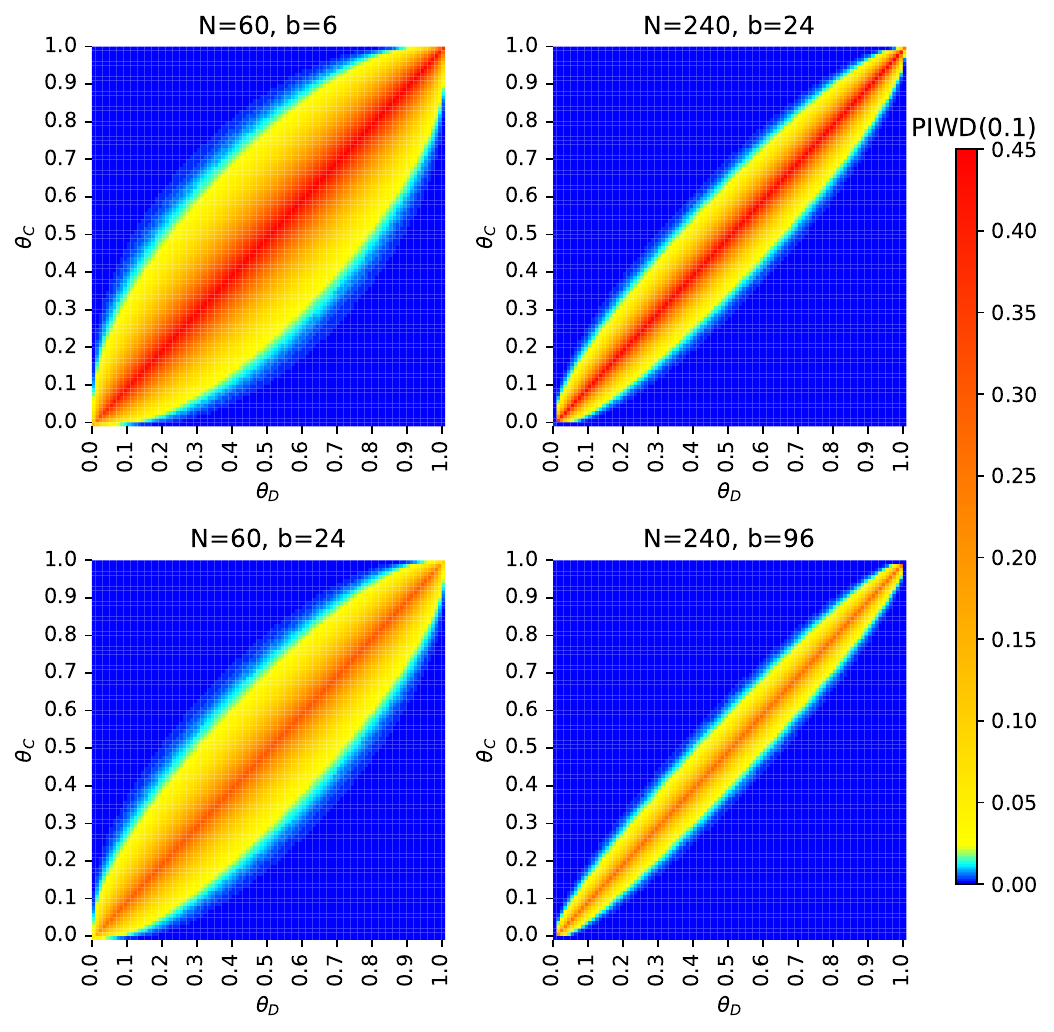}
		\caption{Probability of an imbalance in the wrong direction~(PIWD) across the parameter space for~$\theta_\C,\theta_\D\in \{0.00,0.01,\dots,1.00\}$. burn-in lengths $b\in\{0.1\bar{i}, 0.4\bar{i}\}$ are considered, as well as trial sizes~$\Iend\in\{60,240\}$. For continuity purposes, PIWD(0.1) is set to the maximum value~$0.45$ on the diagonal~$\theta_\C=\theta_{\D}$.}
		\label{fig:S01_heatmap}
	\end{figure}

	Figure~\ref{fig:S01_heatmap} presents a heatmap of values of PIWD(0.1)  across the parameter space, where we restrict to a grid of \hbox{values~$\theta_\C,\theta_\D\in\{0.01,0.02,\dots,1.00\}$} with $\theta_\C\neq\theta_\D$ which gives a clearer picture on where the imbalance occurs.
	\FloatBarrier
	\FloatBarrier

	As also indicated in~\citet{Robertson_Lee_López-Kolkovska_Villar_2023} (who restricted to the case~$\theta_\C=0.25$,~$\theta_\D\geq \theta_\C$ \hbox{and~$\Iend=200$}), Figure~\ref{fig:S01_heatmap} shows that the PIWD(0.1) is highest near the diagonal $\theta_\C=\theta_\D$. The graphs on the bottom row correspond to higher burn-in lengths and show a smaller PIWD(0.1) compared to the top row. 
	Hence, the maximum of PIWD(0.1) shown in Figure~\ref{fig:imbl_epasa} happens around the point~$(1/2 - \delta/2,1/2 + \delta/2)$.
	One of the limitations of the PIWD(0.1) metric is that it is large when the difference in treatments is smallest, and so the cost of allocating to the wrong treatment is lowest, as noted in~\citet{Robertson_Lee_López-Kolkovska_Villar_2023}. Inspection of the numerical results shows that the heatmaps are not symmetrical around the line~$\theta_\C+\theta_\D=1$, which is not noticeable from visual inspection. 
	An explanation could be that the arm with the highest variance switches when comparing a parameter vector with its reflection along the line~$\theta_\C+\theta_\D=1$, hence for one of these scenarios it is more difficult to identify the best arm. For example, comparing the vector~$(0.4,0.5)$ with~$(0.5,0.6)$ the superior arm has the highest variance for the first parameter vector, while for the second it has the smallest variance.
	\FloatBarrier
	\subsection{Bias of treatment effect estimate}\label{sect:bias_results}
	
\begin{figure}[h!]
		
		\centering
		\includegraphics[width=\linewidth]{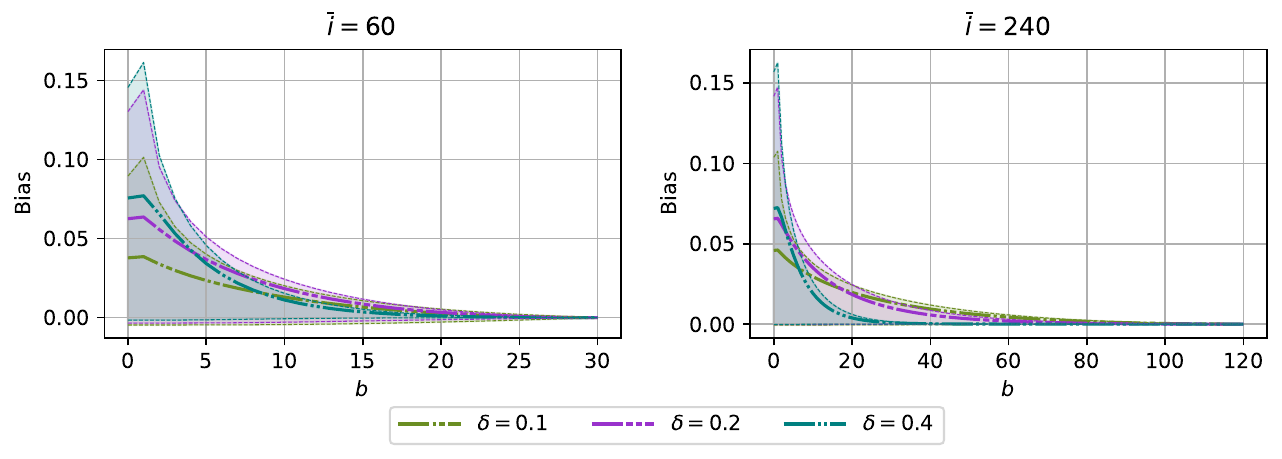}
		\caption{Bias under  treatment effects~$\delta\in \{0.1,0.2,0.4\}$ for the Bayesian response-adaptive randomization design with trial size~$\Iend=60$ and~$\Iend=240$ across different burn-in lengths~$b$.
			The averages for each treatment effect~$\delta$ and fixed burn-in length are represented by thick lines, while the minimum and maximum across the parameter space are presented by thin dotted lines and the ribbons. }
		\label{fig:bias_vs_burnin}
	\end{figure}

	This section investigates the bias of the treatment effect estimator~(Section~\ref{OCscalc}) versus the burn-in length. 
	Figure~\ref{fig:bias_vs_burnin} displays the average and grid-approximated minimum and maximum bias  
	versus
	the burn-in length for treatment \hbox{effects~$\delta\in\{0.1,0.2,0.4\}$.} 
	As one might expect from~\citet{bowden2017unbiased}, the treatment effect estimator largely has a positive bias, while negative values also occur for~$\Iend = 60$ when the developmental treatment has a larger outcome variance than the control treatment~(i.e., when~$\theta_{\D}\leq0.5$). Surprisingly, the maximum and average bias, while starting at higher values for~$b=0$, decreases faster in~$b$ for larger treatment effects~$\delta$. In addition, a lower bias is seen for~$b=0$ than for~$b=1$ which might be due to the case distinction made in the estimator when~$\min_a n_a(\bX_{\Iend})=0$, which decreases the variance of~$\hat{\theta}(\bX_{\Iend})$. Figure~\ref{fig:bias_vs_burnin} shows that, to reduce bias, setting a larger burn-in is better (in case we assume~$b>0$). For a fixed burn-in length~$b$, if we increase ~$\Iend$, the bias tends to be larger on average and in the maximum case, indicating that it might be a good idea to make the burn-in length a function of the trial size~$\Iend$ to make it satisfy the same upper bound. We note that, for a trial setting at hand, bias reduction techniques such as the ones proposed in~\citet{bowden2017unbiased} can be used, which require one to make a (case specific) choice of debiasing technique and bias-variance trade-off.
	
	\FloatBarrier
	\subsection{Recommendation for burn-in} \label{sect:recommend}
	The general takeaway from our analysis is that  while the addition of a burn-in period yields a more balanced type I error rate profile for common non-exact tests (calibration and asymptotic), it cannot guarantee  type I error rate control across the parameter space.
	Non-exact tests applied with insufficient burn-in lengths can exhibit substantial type I error rate inflation (approaching three times the nominal significance level). Therefore, practitioners must exercise caution when using calibrated or asymptotic tests, unless their statistical properties have been thoroughly investigated through simulation or exact methods.  For exploratory settings where strict type I error rate control is not paramount, a practical rule of thumb is to set $b=\Iend/4$ for the BRAR design, which leads to a balanced type I error rate and robust power. If estimation quality is the primary concern, we recommend choosing the smallest burn-in length greater than $b=\Iend/4$ that yields a desired bound on the maximum bias.
	
	In settings where strict type I error control is required (e.g., in confirmatory settings), we recommend considering exact tests. 
	In such settings, type I error rate control is guaranteed, and the burn-in length can be set such that a sufficient power is reached. As we saw that for burn-in lengths at least up to~$\Iend/4$ the CX-S test has higher average power than the UX test, the CX-S test could be preferred in designs that at least target a moderate amount of response-adaptivity. 
	The above guidelines do not consider the PIWD(0.1) due to
	the difficulties with the interpretation of this OC~(as indicated in Section~\ref{sect:participant}).
	We note that if type I error rate and power are not deemed important, it might still be better to use a burn-in, as the maximum average EPASA for our considered trial sizes occurs at small but positive burn-in lengths.
	
	The guidelines above mainly
	focus on the 
	average OCs.
	Since different trials have different priorities,  practitioners may also set threshold values for the OCs which suit their needs, and then investigate which burn-in length satisfies the conditions.
	As we saw that the power and EPASA can have a non-monotonic behaviour in the burn-in length, it is recommended to inspect more burn-in lengths than just~$b=0$ and~$b=\Iend/2$, and inspect at least the values of~$b$ close to these two endpoints. This recommendation is further supported by the optimal burn-in proportions presented in Figure~\ref{fig:optimal_burnin_PPCS} and Figure~\ref{fig:optimal_burnin_Wald},
	where many burn-in lengths are found that maximize power which are strictly between~0 and~$\Iend/2.$
	
	Let us now give an example in a more specific setting. Suppose we want to run a trial with $\bar{i}=240$ participants, hoping to achieve a minimum power of 80\%~under a treatment effect $\delta=0.2$ at a 5\% significance level, while keeping PIWD(0.1) $<5\%$ in~$\Theta_{0.2}$.
	If strict type I error control is required, then we can use exact tests. Figure~\ref{fig:CXUX_PPS} shows that minimum power can be attained using the CX-S test with~$b=63$ or UX test with~$b=68$. If we are lenient with type I error control, we can use the calibrated test with maximum type I error rate $<6\%$. A numerical evaluation then shows that~$b=63$ would be sufficient. For~$\delta=0.2,$ PIWD(0.1) is very close to zero for $b\geq60$, which does not put further restrictions on $b$. Thus we can recommend $b=63$ for the calibrated test, $b=63$ for the CX-S, and~$b=68$ for the UX test as it maximizes the EPASA while fulfilling all the threshold values for the OCs. 
	We would prefer to use the CX-S test in this scenario, which yields the same participant benefit, while it has a stronger type I error rate control than the calibrated test. From this example, we see that even when type I error rate control is not of highest importance, an exact test might give more desirable OCs. As indicated in Figure~\ref{fig:bias_vs_burnin}, this burn-in value yields a maximum bias~(over~$\delta\in\{0.1,0.2,0.4\}$) of around 0.51\%, which might be acceptable in practical settings.
	
	\subsection{Sensitivity analysis to the choice of prior}\label{sect:sensitivity_analysis}
	This paper considers BRAR using the Beta(1,1) prior.  This section provides a sensitivity analysis of our findings with respect to this prior choice. Under the Beta(1,1) prior our main findings are~(1) standard~(non-exact) tests show
	substantial type I error rate inflation which can be reduced but not fully mitigated through a larger burn-in length;~(2) exact tests can demonstrate superior power over standard
	tests, with the conditional~(unconditional) exact test showing higher power for smaller~(larger) burn-in lengths;~(3) 
	the choice of test statistic affects  power and type I error rate;~(4) statistical bias decreases more quickly in the burn-in length for large
	treatment effects and is higher for larger trial sizes;~(5) power and EPASA are not always maximized at the largest and smallest burn-in lengths, respectively. 
	
	Table~\ref{tab:T1E_sens_prior} shows average and maximum type I error rates for the calibrated and exact PPCS tests, as well as the asymptotic Wald test for different priors for both treatment arms~(both used for BRAR and testing) and trial sizes~$\Iend\in\{60,240\}$. The average and maximum type I error rate for the calibrated PPCS test under the Beta(.01, .01) prior substantially increase in comparison to the uniform prior (which likely stems from its heavy prior mass near 0 and 1, together with the reasoning given in the third paragraph of Section \ref{typeIerrorcurves}), showing that the type I error rate under the BRAR design with the calibrated PPCS test is sensitive to the prior. 
	As the maximum type I error rate remains substantially larger than 6\%, finding~(1) no longer holds in this setting. The other priors show a lower degree of additional type I error inflation: Beta(0.5, 0.5) and Beta(1.4,0.6) result in a maximum type I error rate above 7\%  for~$\Iend/4$ under~$\Iend=60$. This increase in type I error rate inflation may be due to to a higher probability mass given to large common success rates under the Beta(.01, .01), Beta(0.5, 0.5) and Beta(1.4,0.6) priors. 
	Under the asymptotic Wald test, using the prior only to determine the allocation probabilities, the differences are smaller and the rule~$b\geq\Iend/4$  shows maximum type I error rates below~$7\%$, agreeing with finding~(1). For all priors except~Beta(.01, .01) we find agreement with finding~(2): the CX-S test is less~(more) conservative  than the UX test on average for smaller~(larger) burn-ins. The conservativeness of the UX test for Beta(.01, .01) comes from correcting the substantial type I error rate inflation under this prior. The type I error rates vary greatly when comparing the calibrated PPCS and asymptotic Wald tests, agreeing with finding~(3). The maximum type I error rates are most stable for the uniform prior, and the average type I error rates under the CX-S PPCS test are robust to the chosen prior.

    \begin{table}[h!]
   \centering
    \caption{Average and maximum type I error rates (percentage points) under Bayesian response-adaptive randomization for different  prior configurations, trial sizes~$\Iend$, and burn-in lengths~$b$. Type I error rate inflation above the~6\% level is indicated by bold~(and a color gradient from brown to red). }
    \hspace{-5mm}\scriptsize\def\arraystretch{1.2}\setlength{\tabcolsep}{2pt}\begin{tabular}{llcccccccccccc}
\toprule
 &  & \multicolumn{6}{c}{$\Iend = $ 60} & \multicolumn{6}{c}{$\Iend = $ 240} \\
 &  & AFP-C & MFP-C & AFP-CX & AFP-UX & AFP-W & MFP-W & AFP-C & MFP-C & AFP-CX & AFP-UX & AFP-W & MFP-W \\
Prior & $b$ &  &  &  &  &  &  &  &  &  &  &  &  \\
\midrule
\multirow[t]{5}{*}{Beta(1.0, 1.0)} & 0 & 5.59 & \textcolor{red}{\textbf{14.53}} & 4.00 & 1.02 & \textcolor{brown}{\textbf{6.38}} & \textcolor{red}{\textbf{10.31}} & 5.06 & \textcolor{red}{\textbf{13.31}} & 4.75 & 0.77 & \textcolor{orange}{\textbf{9.56}} & \textcolor{red}{\textbf{12.22}} \\
 & $\Iend/10$ & 4.34 & \textcolor{orange}{\textbf{8.59}} & 4.17 & 2.25 & 5.08 & \textcolor{orange}{\textbf{8.21}} & 4.61 & \textcolor{brown}{\textbf{7.82}} & 4.85 & 2.67 & \textcolor{brown}{\textbf{6.45}} & \textcolor{orange}{\textbf{8.32}} \\
 & $\Iend/4$ & 4.18 & \textcolor{brown}{\textbf{6.36}} & 4.02 & 3.10 & 4.04 & 5.92 & 4.70 & \textcolor{brown}{\textbf{6.30}} & 4.81 & 3.75 & 4.88 & 5.83 \\
 & $\Iend \cdot 4/10$ & 4.02 & 5.15 & 3.15 & 3.91 & 3.78 & 5.17 & 4.67 & 5.17 & 4.53 & 4.32 & 4.51 & 5.11 \\
 & $\Iend/2$ & 3.36 & 4.69 & 1.94 & 3.36 & 3.88 & 5.19 & 4.98 & 5.66 & 3.15 & 4.43 & 4.41 & 5.18 \\
\cline{1-14}
\multirow[t]{5}{*}{Beta(.01, .01)} & 0 & 3.13 & 4.97 & 3.18 & 3.13 & 4.22 & \textcolor{orange}{\textbf{9.16}} & \textcolor{red}{\textbf{13.83}} & \textcolor{red}{\textbf{60.65}} & 4.55 & 1.72 & \textcolor{brown}{\textbf{7.82}} & \textcolor{red}{\textbf{12.30}} \\
 & $\Iend/10$ & \textcolor{red}{\textbf{18.18}} & \textcolor{red}{\textbf{72.28}} & 3.73 & 1.18 & 4.66 & \textcolor{orange}{\textbf{8.22}} & \textcolor{orange}{\textbf{9.03}} & \textcolor{red}{\textbf{73.92}} & 4.76 & 0.43 & \textcolor{brown}{\textbf{6.28}} & \textcolor{orange}{\textbf{8.41}} \\
 & $\Iend/4$ & \textcolor{red}{\textbf{12.65}} & \textcolor{red}{\textbf{68.68}} & 3.77 & 0.96 & 3.94 & 5.91 & \textcolor{brown}{\textbf{7.31}} & \textcolor{red}{\textbf{68.13}} & 4.71 & 0.28 & 4.86 & 5.82 \\
 & $\Iend \cdot 4/10$ & \textcolor{red}{\textbf{10.51}} & \textcolor{red}{\textbf{56.41}} & 3.02 & 0.90 & 3.80 & 5.17 & \textcolor{brown}{\textbf{6.63}} & \textcolor{red}{\textbf{51.99}} & 4.44 & 0.35 & 4.51 & 5.12 \\
 & $\Iend/2$ & \textcolor{orange}{\textbf{9.93}} & \textcolor{red}{\textbf{49.59}} & 1.94 & 0.80 & 3.88 & 5.19 & \textcolor{brown}{\textbf{6.92}} & \textcolor{red}{\textbf{42.05}} & 3.15 & 0.34 & 4.41 & 5.18 \\
\cline{1-14}
\multirow[t]{5}{*}{Beta(0.5, 0.5)} & 0 & 4.63 & \textcolor{red}{\textbf{10.15}} & 3.63 & 1.61 & \textcolor{brown}{\textbf{6.17}} & \textcolor{red}{\textbf{11.28}} & 4.01 & \textcolor{brown}{\textbf{6.84}} & 4.56 & 2.17 & \textcolor{orange}{\textbf{9.99}} & \textcolor{red}{\textbf{13.63}} \\
 & $\Iend/10$ & 3.87 & 5.73 & 4.10 & 3.28 & 4.95 & \textcolor{orange}{\textbf{8.51}} & 4.78 & \textcolor{brown}{\textbf{6.11}} & 4.82 & 3.82 & \textcolor{brown}{\textbf{6.40}} & \textcolor{orange}{\textbf{8.41}} \\
 & $\Iend/4$ & 4.52 & \textcolor{brown}{\textbf{7.14}} & 3.89 & 3.16 & 4.00 & 5.92 & 5.01 & \textcolor{brown}{\textbf{7.26}} & 4.77 & 3.51 & 4.87 & 5.83 \\
 & $\Iend \cdot 4/10$ & 4.56 & 5.81 & 3.02 & 3.69 & 3.80 & 5.17 & 4.94 & 5.69 & 4.52 & 3.18 & 4.51 & 5.12 \\
 & $\Iend/2$ & 4.18 & 5.53 & 1.94 & 3.95 & 3.88 & 5.19 & 5.32 & \textcolor{brown}{\textbf{6.62}} & 3.15 & 4.60 & 4.41 & 5.18 \\
\cline{1-14}
\multirow[t]{5}{*}{Beta(1.4, 0.6)} & 0 & 5.30 & \textcolor{red}{\textbf{14.34}} & 4.30 & 1.29 & \textcolor{brown}{\textbf{6.18}} & \textcolor{orange}{\textbf{9.34}} & 4.72 & \textcolor{red}{\textbf{11.03}} & 4.87 & 1.32 & \textcolor{orange}{\textbf{8.65}} & \textcolor{red}{\textbf{10.89}} \\
 & $\Iend/10$ & 4.54 & \textcolor{orange}{\textbf{9.95}} & 4.36 & 1.90 & 5.17 & \textcolor{orange}{\textbf{8.05}} & 4.43 & \textcolor{brown}{\textbf{7.59}} & 4.88 & 2.70 & \textcolor{brown}{\textbf{6.49}} & \textcolor{orange}{\textbf{8.22}} \\
 & $\Iend/4$ & 4.33 & \textcolor{brown}{\textbf{7.99}} & 4.14 & 2.36 & 4.07 & 5.94 & 4.62 & \textcolor{brown}{\textbf{7.73}} & 4.81 & 3.05 & 4.90 & 5.84 \\
 & $\Iend \cdot 4/10$ & 3.92 & 5.52 & 3.11 & 3.26 & 3.77 & 5.17 & 4.67 & 5.87 & 4.55 & 3.28 & 4.51 & 5.12 \\
 & $\Iend/2$ & 3.57 & 4.69 & 1.94 & 3.57 & 3.88 & 5.19 & 5.05 & \textcolor{brown}{\textbf{6.62}} & 3.15 & 4.50 & 4.41 & 5.18 \\
\cline{1-14}
\multirow[t]{5}{*}{Beta(0.6, 1.4)} & 0 & \textcolor{brown}{\textbf{7.05}} & \textcolor{red}{\textbf{34.44}} & 3.36 & 0.81 & 5.60 & \textcolor{red}{\textbf{10.11}} & \textcolor{brown}{\textbf{6.14}} & \textcolor{red}{\textbf{37.68}} & 4.16 & 0.44 & \textcolor{red}{\textbf{10.60}} & \textcolor{red}{\textbf{14.41}} \\
 & $\Iend/10$ & 3.93 & \textcolor{brown}{\textbf{6.87}} & 3.91 & 2.64 & 4.86 & \textcolor{orange}{\textbf{8.24}} & 4.91 & \textcolor{orange}{\textbf{8.21}} & 4.82 & 2.69 & \textcolor{brown}{\textbf{6.33}} & \textcolor{orange}{\textbf{8.34}} \\
 & $\Iend/4$ & 4.32 & 5.72 & 3.86 & 3.80 & 3.99 & 5.90 & 4.87 & 5.96 & 4.80 & 4.02 & 4.86 & 5.82 \\
 & $\Iend \cdot 4/10$ & 4.18 & 4.99 & 3.18 & 4.12 & 3.79 & 5.18 & 4.80 & 5.11 & 4.52 & 4.70 & 4.51 & 5.12 \\
 & $\Iend/2$ & 3.57 & 4.69 & 1.94 & 3.57 & 3.88 & 5.19 & 5.05 & 6.62 & 3.15 & 4.50 & 4.41 & 5.18 \\
\cline{1-14}
\bottomrule
\end{tabular}
\\
    AFP-C: average type I error~(false positive) rate for the calibrated posterior probability of control superiority~(PPCS) test;  MFP-C: maximum type I error~(false positive)  rate for the calibrated PPCS test; AFP-CX: average type I error~(false positive)  rate for the conditional exact PPCS test; AFP-UX: average type I error~(false positive)  rate for the unconditional exact PPCS test; AFP-W: average type I error~(false positive) rate for the asymptotic Wald test;  MFP-W: maximum type I error~(false positive)  rate for the  asymptotic Wald test. Beta($\alpha,\beta$) denotes the beta distribution with parameters~$\alpha,\beta$.
    \label{tab:T1E_sens_prior}
\end{table}
\FloatBarrier
Table~\ref{tab:tab_prior_sens} provides the minimum power of the calibrated PPCS test,  exact PPCS tests, and asymptotic  Wald test, as well as average EPASA and bias  under a BRAR design for different  prior configurations, trial sizes, treatment effects, and burn-in lengths. Comparing the minimum power values for~$\Iend=60$,  the minimum power of the CX-S PPCS test is often higher than that of the other tests, except for Beta(.01, .01)~(where substantial type I error inflation was found), agreeing with finding~(2). Power improvements for~$
	\Iend=240$ under the CX-S test are only seen across PPCS tests, with the asymptotic Wald test showing higher power, agreeing with finding~(3). Looking at the bias, noticing that~$\Iend\cdot 4/10 = 24$ for~$\Iend=60$ and~$\Iend/10=24$ when~$\Iend=240$, we find agreement with finding~(4) across prior configurations. 
	In agreement with finding~(5), the average EPASA attains a maximum for~$\Iend/10$ for the Beta(.01,.01) prior for~$\delta=0.2$. In general, differences between OCs decrease for larger burn-in lengths. The prior~Beta(1.4, 0.6) often improves power and decreases bias in comparison to the uniform prior~(at the cost of EPASA), which might be because the BRAR procedure is less sensitive to early successes in this setting.

\begin{table}[h!]
    \centering
\scriptsize\caption{Average operating characteristics (percentage points) under Bayesian response-adaptive randomization for different  prior configurations, trial sizes~$\Iend$, treatment effects~$\delta$, and burn-in lengths~$b$. Changes in OCs for each non-uniform prior with the uniform~(Beta(1.0, 1.0)) prior are displayed, with positive changes overlined~(orange) and negative changes underlined~(blue).   }
    \def\arraystretch{1.15}\setlength{\tabcolsep}{2pt}\begin{tabular}{lllrrrrrrrrrrrr}
\toprule
 &  &  & \multicolumn{6}{c}{$\Iend = $ 60} & \multicolumn{6}{c}{$\Iend = $ 240} \\
 &  &  & MP-C & MP-CX & MP-UX & MP-W & EPASA & Bias & MP-C & MP-CX & MP-UX & MP-W & EPASA & Bias \\
Prior & $\delta$ & $b$ &  &  &  &  &  &  &  &  &  &  &  &  \\
\midrule
\multirow[t]{9}{*}{Beta(1.0, 1.0)} & \multirow[t]{3}{*}{0.10} & 0 & 0.18 & 6.23 & 0.00 & 0.51 & 65.56 & 3.78 & 4.22 & 11.37 & 0.10 & 28.33 & 78.35 & 4.58 \\
 &  & $\Iend/10$ & 0.16 & 6.89 & 0.03 & 0.55 & 64.82 & 2.09 & 18.98 & 18.96 & 6.06 & 26.84 & 76.43 & 1.70 \\
 &  & $\Iend\cdot 4/10$ & 11.10 & 9.10 & 7.90 & 11.86 & 54.72 & 0.10 & 33.02 & 32.04 & 32.14 & 33.37 & 57.85 & 0.06 \\
\cline{2-15}
 & \multirow[t]{3}{*}{0.20} & 0 & 3.98 & 13.05 & 0.12 & 13.64 & 76.72 & 6.26 & 30.71 & 31.42 & 1.07 & 53.49 & 89.86 & 6.55 \\
 &  & $\Iend/10$ & 4.61 & 16.63 & 0.94 & 16.37 & 75.19 & 3.20 & 58.52 & 57.64 & 46.30 & 68.73 & 86.12 & 1.44 \\
 &  & $\Iend\cdot 4/10$ & 31.89 & 28.03 & 31.07 & 33.63 & 57.67 & 0.11 & 86.51 & 85.85 & 85.97 & 86.69 & 59.75 & 0.01 \\
\cline{2-15}
 & \multirow[t]{3}{*}{0.40} & 0 & 39.65 & 37.44 & 4.16 & 61.39 & 88.61 & 7.57 & 79.86 & 70.93 & 10.23 & 76.24 & 96.21 & 7.20 \\
 &  & $\Iend/10$ & 61.39 & 55.97 & 36.48 & 70.66 & 85.61 & 2.73 & 99.08 & 98.89 & 97.87 & 99.48 & 89.80 & 0.21 \\
 &  & $\Iend\cdot 4/10$ & 87.20 & 84.60 & 86.85 & 88.33 & 59.78 & 0.03 & 100.00 & 100.00 & 100.00 & 100.00 & 59.94 & 0.00 \\
\cline{1-15} \cline{2-15}
\multirow[t]{9}{*}{Beta(.01, .01)} & \multirow[t]{3}{*}{0.10} & 0 & $\overline{\textcolor{darkorange}{\bf2.72}}$ & $\uline{\textcolor{royalblue}{\bf2.49}}$ & $\overline{\textcolor{darkorange}{\bf0.07}}$ & $\uline{\textcolor{royalblue}{\bf0.25}}$ & $\uline{\textcolor{royalblue}{\bf65.07}}$ & $\uline{\textcolor{royalblue}{\bf3.40}}$ & $\overline{\textcolor{darkorange}{\bf10.94}}$ & $\uline{\textcolor{royalblue}{\bf7.59}}$ & $\overline{\textcolor{darkorange}{\bf0.76}}$ & $\uline{\textcolor{royalblue}{\bf7.84}}$ & $\uline{\textcolor{royalblue}{\bf76.13}}$ & $\uline{\textcolor{royalblue}{\bf3.85}}$ \\
 &  & $\Iend/10$ & $\overline{\textcolor{darkorange}{\bf7.61}}$ & $\uline{\textcolor{royalblue}{\bf5.21}}$ & $\overline{\textcolor{darkorange}{\bf0.05}}$ & $\uline{\textcolor{royalblue}{\bf0.18}}$ & $\overline{\textcolor{darkorange}{\bf65.82}}$ & $\overline{\textcolor{darkorange}{\bf2.14}}$ & $\uline{\textcolor{royalblue}{\bf18.66}}$ & $\uline{\textcolor{royalblue}{\bf18.50}}$ & $\uline{\textcolor{royalblue}{\bf0.61}}$ & $\uline{\textcolor{royalblue}{\bf26.75}}$ & $\overline{\textcolor{darkorange}{\bf76.73}}$ & $\overline{\textcolor{darkorange}{\bf1.72}}$ \\
 &  & $\Iend\cdot 4/10$ & $\uline{\textcolor{royalblue}{\bf11.08}}$ & $\uline{\textcolor{royalblue}{\bf8.95}}$ & $\uline{\textcolor{royalblue}{\bf0.66}}$ & $\uline{\textcolor{royalblue}{\bf11.84}}$ & $\overline{\textcolor{darkorange}{\bf54.94}}$ & 0.10 & $\overline{\textcolor{darkorange}{\bf33.05}}$ & $\uline{\textcolor{royalblue}{\bf32.03}}$ & $\uline{\textcolor{royalblue}{\bf4.16}}$ & $\uline{\textcolor{royalblue}{\bf33.32}}$ & $\overline{\textcolor{darkorange}{\bf57.89}}$ & 0.06 \\
\cline{2-15}
 & \multirow[t]{3}{*}{0.20} & 0 & $\overline{\textcolor{darkorange}{\bf5.60}}$ & $\uline{\textcolor{royalblue}{\bf5.40}}$ & $\overline{\textcolor{darkorange}{\bf1.77}}$ & $\uline{\textcolor{royalblue}{\bf3.66}}$ & $\uline{\textcolor{royalblue}{\bf75.27}}$ & $\uline{\textcolor{royalblue}{\bf5.60}}$ & $\overline{\textcolor{darkorange}{\bf30.87}}$ & $\uline{\textcolor{royalblue}{\bf22.71}}$ & $\overline{\textcolor{darkorange}{\bf4.54}}$ & $\uline{\textcolor{royalblue}{\bf30.16}}$ & $\uline{\textcolor{royalblue}{\bf87.55}}$ & $\uline{\textcolor{royalblue}{\bf5.39}}$ \\
 &  & $\Iend/10$ & $\overline{\textcolor{darkorange}{\bf17.31}}$ & $\uline{\textcolor{royalblue}{\bf13.24}}$ & $\overline{\textcolor{darkorange}{\bf0.97}}$ & $\uline{\textcolor{royalblue}{\bf4.25}}$ & $\overline{\textcolor{darkorange}{\bf76.15}}$ & $\uline{\textcolor{royalblue}{\bf3.19}}$ & $\uline{\textcolor{royalblue}{\bf56.55}}$ & $\uline{\textcolor{royalblue}{\bf56.71}}$ & $\uline{\textcolor{royalblue}{\bf5.66}}$ & $\uline{\textcolor{royalblue}{\bf68.49}}$ & $\overline{\textcolor{darkorange}{\bf86.23}}$ & $\uline{\textcolor{royalblue}{\bf1.43}}$ \\
 &  & $\Iend\cdot 4/10$ & $\uline{\textcolor{royalblue}{\bf31.83}}$ & $\uline{\textcolor{royalblue}{\bf27.75}}$ & $\uline{\textcolor{royalblue}{\bf4.18}}$ & $\uline{\textcolor{royalblue}{\bf33.61}}$ & $\overline{\textcolor{darkorange}{\bf57.82}}$ & 0.11 & 86.51 & $\overline{\textcolor{darkorange}{\bf85.86}}$ & $\uline{\textcolor{royalblue}{\bf42.01}}$ & $\uline{\textcolor{royalblue}{\bf86.66}}$ & $\overline{\textcolor{darkorange}{\bf59.79}}$ & 0.01 \\
\cline{2-15}
 & \multirow[t]{3}{*}{0.40} & 0 & $\uline{\textcolor{royalblue}{\bf17.65}}$ & $\uline{\textcolor{royalblue}{\bf11.64}}$ & $\overline{\textcolor{darkorange}{\bf13.78}}$ & $\uline{\textcolor{royalblue}{\bf20.32}}$ & $\uline{\textcolor{royalblue}{\bf87.07}}$ & $\uline{\textcolor{royalblue}{\bf6.69}}$ & $\uline{\textcolor{royalblue}{\bf67.80}}$ & $\uline{\textcolor{royalblue}{\bf66.67}}$ & $\overline{\textcolor{darkorange}{\bf16.04}}$ & $\uline{\textcolor{royalblue}{\bf64.47}}$ & $\uline{\textcolor{royalblue}{\bf94.82}}$ & $\uline{\textcolor{royalblue}{\bf4.39}}$ \\
 &  & $\Iend/10$ & $\uline{\textcolor{royalblue}{\bf54.77}}$ & $\uline{\textcolor{royalblue}{\bf49.08}}$ & $\uline{\textcolor{royalblue}{\bf7.73}}$ & $\uline{\textcolor{royalblue}{\bf68.39}}$ & $\overline{\textcolor{darkorange}{\bf85.95}}$ & $\uline{\textcolor{royalblue}{\bf2.36}}$ & $\uline{\textcolor{royalblue}{\bf98.70}}$ & $\uline{\textcolor{royalblue}{\bf98.84}}$ & $\uline{\textcolor{royalblue}{\bf64.54}}$ & $\overline{\textcolor{darkorange}{\bf99.51}}$ & $\uline{\textcolor{royalblue}{\bf89.69}}$ & $\uline{\textcolor{royalblue}{\bf0.16}}$ \\
 &  & $\Iend\cdot 4/10$ & $\uline{\textcolor{royalblue}{\bf87.16}}$ & $\uline{\textcolor{royalblue}{\bf84.39}}$ & $\uline{\textcolor{royalblue}{\bf43.78}}$ & $\uline{\textcolor{royalblue}{\bf88.30}}$ & 59.77 & $\uline{\textcolor{royalblue}{\bf0.02}}$ & 100.00 & 100.00 & $\uline{\textcolor{royalblue}{\bf99.87}}$ & 100.00 & $\overline{\textcolor{darkorange}{\bf59.99}}$ & 0.00 \\
\cline{1-15} \cline{2-15}
\multirow[t]{9}{*}{Beta(0.5, 0.5)} & \multirow[t]{3}{*}{0.10} & 0 & $\overline{\textcolor{darkorange}{\bf0.54}}$ & $\uline{\textcolor{royalblue}{\bf5.37}}$ & $\overline{\textcolor{darkorange}{\bf0.03}}$ & $\uline{\textcolor{royalblue}{\bf0.25}}$ & $\overline{\textcolor{darkorange}{\bf66.14}}$ & $\overline{\textcolor{darkorange}{\bf4.28}}$ & $\overline{\textcolor{darkorange}{\bf5.47}}$ & $\uline{\textcolor{royalblue}{\bf9.86}}$ & $\overline{\textcolor{darkorange}{\bf1.32}}$ & $\uline{\textcolor{royalblue}{\bf15.81}}$ & $\overline{\textcolor{darkorange}{\bf78.48}}$ & $\overline{\textcolor{darkorange}{\bf4.88}}$ \\
 &  & $\Iend/10$ & $\overline{\textcolor{darkorange}{\bf0.53}}$ & $\uline{\textcolor{royalblue}{\bf5.84}}$ & $\overline{\textcolor{darkorange}{\bf0.04}}$ & $\uline{\textcolor{royalblue}{\bf0.24}}$ & $\overline{\textcolor{darkorange}{\bf65.39}}$ & $\overline{\textcolor{darkorange}{\bf2.22}}$ & $\overline{\textcolor{darkorange}{\bf19.11}}$ & $\uline{\textcolor{royalblue}{\bf18.72}}$ & $\overline{\textcolor{darkorange}{\bf16.38}}$ & $\uline{\textcolor{royalblue}{\bf26.82}}$ & $\overline{\textcolor{darkorange}{\bf76.64}}$ & $\overline{\textcolor{darkorange}{\bf1.73}}$ \\
 &  & $\Iend\cdot 4/10$ & $\uline{\textcolor{royalblue}{\bf11.09}}$ & $\uline{\textcolor{royalblue}{\bf9.03}}$ & $\uline{\textcolor{royalblue}{\bf7.15}}$ & $\uline{\textcolor{royalblue}{\bf11.85}}$ & $\overline{\textcolor{darkorange}{\bf54.83}}$ & 0.10 & $\overline{\textcolor{darkorange}{\bf33.06}}$ & 32.04 & $\uline{\textcolor{royalblue}{\bf26.69}}$ & $\uline{\textcolor{royalblue}{\bf33.33}}$ & $\overline{\textcolor{darkorange}{\bf57.88}}$ & 0.06 \\
\cline{2-15}
 & \multirow[t]{3}{*}{0.20} & 0 & $\overline{\textcolor{darkorange}{\bf6.12}}$ & $\uline{\textcolor{royalblue}{\bf11.21}}$ & $\overline{\textcolor{darkorange}{\bf1.29}}$ & $\uline{\textcolor{royalblue}{\bf5.58}}$ & $\overline{\textcolor{darkorange}{\bf77.42}}$ & $\overline{\textcolor{darkorange}{\bf7.10}}$ & $\uline{\textcolor{royalblue}{\bf25.19}}$ & $\uline{\textcolor{royalblue}{\bf26.86}}$ & $\overline{\textcolor{darkorange}{\bf5.47}}$ & $\uline{\textcolor{royalblue}{\bf51.78}}$ & $\overline{\textcolor{darkorange}{\bf90.01}}$ & $\overline{\textcolor{darkorange}{\bf7.10}}$ \\
 &  & $\Iend/10$ & $\overline{\textcolor{darkorange}{\bf7.44}}$ & $\uline{\textcolor{royalblue}{\bf14.48}}$ & $\overline{\textcolor{darkorange}{\bf2.20}}$ & $\uline{\textcolor{royalblue}{\bf7.19}}$ & $\overline{\textcolor{darkorange}{\bf75.84}}$ & $\overline{\textcolor{darkorange}{\bf3.38}}$ & $\uline{\textcolor{royalblue}{\bf57.76}}$ & $\uline{\textcolor{royalblue}{\bf57.21}}$ & $\overline{\textcolor{darkorange}{\bf52.84}}$ & $\uline{\textcolor{royalblue}{\bf68.64}}$ & $\overline{\textcolor{darkorange}{\bf86.23}}$ & $\overline{\textcolor{darkorange}{\bf1.45}}$ \\
 &  & $\Iend\cdot 4/10$ & $\uline{\textcolor{royalblue}{\bf31.86}}$ & $\uline{\textcolor{royalblue}{\bf27.89}}$ & $\uline{\textcolor{royalblue}{\bf29.58}}$ & $\uline{\textcolor{royalblue}{\bf33.62}}$ & $\overline{\textcolor{darkorange}{\bf57.76}}$ & 0.11 & $\overline{\textcolor{darkorange}{\bf86.52}}$ & $\overline{\textcolor{darkorange}{\bf85.87}}$ & $\uline{\textcolor{royalblue}{\bf82.10}}$ & $\uline{\textcolor{royalblue}{\bf86.67}}$ & $\overline{\textcolor{darkorange}{\bf59.80}}$ & 0.01 \\
\cline{2-15}
 & \multirow[t]{3}{*}{0.40} & 0 & $\uline{\textcolor{royalblue}{\bf37.87}}$ & $\uline{\textcolor{royalblue}{\bf31.16}}$ & $\overline{\textcolor{darkorange}{\bf12.11}}$ & $\uline{\textcolor{royalblue}{\bf45.99}}$ & $\overline{\textcolor{darkorange}{\bf89.27}}$ & $\overline{\textcolor{darkorange}{\bf8.67}}$ & $\uline{\textcolor{royalblue}{\bf70.16}}$ & $\uline{\textcolor{royalblue}{\bf64.84}}$ & $\overline{\textcolor{darkorange}{\bf17.68}}$ & $\overline{\textcolor{darkorange}{\bf79.41}}$ & $\overline{\textcolor{darkorange}{\bf96.37}}$ & $\overline{\textcolor{darkorange}{\bf7.87}}$ \\
 &  & $\Iend/10$ & $\uline{\textcolor{royalblue}{\bf57.43}}$ & $\uline{\textcolor{royalblue}{\bf54.19}}$ & $\overline{\textcolor{darkorange}{\bf52.21}}$ & $\uline{\textcolor{royalblue}{\bf70.10}}$ & $\overline{\textcolor{darkorange}{\bf86.02}}$ & $\overline{\textcolor{darkorange}{\bf2.80}}$ & $\uline{\textcolor{royalblue}{\bf98.92}}$ & $\uline{\textcolor{royalblue}{\bf98.87}}$ & $\overline{\textcolor{darkorange}{\bf98.22}}$ & $\overline{\textcolor{darkorange}{\bf99.49}}$ & $\overline{\textcolor{darkorange}{\bf89.81}}$ & 0.20 \\
 &  & $\Iend\cdot 4/10$ & $\uline{\textcolor{royalblue}{\bf87.18}}$ & $\uline{\textcolor{royalblue}{\bf84.51}}$ & $\uline{\textcolor{royalblue}{\bf85.78}}$ & $\uline{\textcolor{royalblue}{\bf88.32}}$ & $\overline{\textcolor{darkorange}{\bf59.79}}$ & 0.03 & 100.00 & 100.00 & 100.00 & 100.00 & $\overline{\textcolor{darkorange}{\bf60.00}}$ & 0.00 \\
\cline{1-15} \cline{2-15}
\multirow[t]{9}{*}{Beta(1.4, 0.6)} & \multirow[t]{3}{*}{0.10} & 0 & $\overline{\textcolor{darkorange}{\bf0.24}}$ & $\overline{\textcolor{darkorange}{\bf7.34}}$ & 0.01 & $\overline{\textcolor{darkorange}{\bf0.96}}$ & $\uline{\textcolor{royalblue}{\bf64.93}}$ & $\uline{\textcolor{royalblue}{\bf3.21}}$ & $\overline{\textcolor{darkorange}{\bf6.31}}$ & $\overline{\textcolor{darkorange}{\bf13.66}}$ & $\overline{\textcolor{darkorange}{\bf0.42}}$ & $\uline{\textcolor{royalblue}{\bf25.20}}$ & $\uline{\textcolor{royalblue}{\bf77.77}}$ & $\uline{\textcolor{royalblue}{\bf3.81}}$ \\
 &  & $\Iend/10$ & $\overline{\textcolor{darkorange}{\bf0.19}}$ & $\overline{\textcolor{darkorange}{\bf7.67}}$ & $\uline{\textcolor{royalblue}{\bf0.02}}$ & $\overline{\textcolor{darkorange}{\bf1.02}}$ & $\uline{\textcolor{royalblue}{\bf64.23}}$ & $\uline{\textcolor{royalblue}{\bf1.96}}$ & $\uline{\textcolor{royalblue}{\bf15.40}}$ & $\overline{\textcolor{darkorange}{\bf19.49}}$ & $\uline{\textcolor{royalblue}{\bf5.36}}$ & $\overline{\textcolor{darkorange}{\bf27.48}}$ & $\uline{\textcolor{royalblue}{\bf75.94}}$ & $\uline{\textcolor{royalblue}{\bf1.66}}$ \\
 &  & $\Iend\cdot 4/10$ & $\uline{\textcolor{royalblue}{\bf6.37}}$ & $\overline{\textcolor{darkorange}{\bf9.12}}$ & $\uline{\textcolor{royalblue}{\bf6.01}}$ & 11.86 & $\uline{\textcolor{royalblue}{\bf54.69}}$ & 0.09 & $\overline{\textcolor{darkorange}{\bf33.09}}$ & $\overline{\textcolor{darkorange}{\bf32.06}}$ & $\uline{\textcolor{royalblue}{\bf27.72}}$ & $\uline{\textcolor{royalblue}{\bf33.34}}$ & 57.84 & 0.06 \\
\cline{2-15}
 & \multirow[t]{3}{*}{0.20} & 0 & $\overline{\textcolor{darkorange}{\bf5.93}}$ & $\overline{\textcolor{darkorange}{\bf15.82}}$ & $\overline{\textcolor{darkorange}{\bf0.35}}$ & $\overline{\textcolor{darkorange}{\bf23.80}}$ & $\uline{\textcolor{royalblue}{\bf75.68}}$ & $\uline{\textcolor{royalblue}{\bf5.28}}$ & $\overline{\textcolor{darkorange}{\bf42.38}}$ & $\overline{\textcolor{darkorange}{\bf39.34}}$ & $\overline{\textcolor{darkorange}{\bf5.56}}$ & $\uline{\textcolor{royalblue}{\bf50.93}}$ & $\uline{\textcolor{royalblue}{\bf89.17}}$ & $\uline{\textcolor{royalblue}{\bf5.44}}$ \\
 &  & $\Iend/10$ & $\uline{\textcolor{royalblue}{\bf4.57}}$ & $\overline{\textcolor{darkorange}{\bf18.37}}$ & $\uline{\textcolor{royalblue}{\bf0.80}}$ & $\overline{\textcolor{darkorange}{\bf26.06}}$ & $\uline{\textcolor{royalblue}{\bf74.35}}$ & $\uline{\textcolor{royalblue}{\bf3.04}}$ & $\overline{\textcolor{darkorange}{\bf59.79}}$ & $\overline{\textcolor{darkorange}{\bf58.90}}$ & $\overline{\textcolor{darkorange}{\bf48.49}}$ & $\overline{\textcolor{darkorange}{\bf70.09}}$ & $\uline{\textcolor{royalblue}{\bf85.82}}$ & $\overline{\textcolor{darkorange}{\bf1.47}}$ \\
 &  & $\Iend\cdot 4/10$ & $\overline{\textcolor{darkorange}{\bf32.06}}$ & $\overline{\textcolor{darkorange}{\bf28.06}}$ & $\uline{\textcolor{royalblue}{\bf28.07}}$ & 33.63 & $\uline{\textcolor{royalblue}{\bf57.64}}$ & 0.10 & $\overline{\textcolor{darkorange}{\bf86.57}}$ & $\overline{\textcolor{darkorange}{\bf85.89}}$ & $\uline{\textcolor{royalblue}{\bf82.89}}$ & $\uline{\textcolor{royalblue}{\bf86.68}}$ & $\overline{\textcolor{darkorange}{\bf59.79}}$ & 0.01 \\
\cline{2-15}
 & \multirow[t]{3}{*}{0.40} & 0 & $\overline{\textcolor{darkorange}{\bf49.96}}$ & $\overline{\textcolor{darkorange}{\bf46.76}}$ & $\overline{\textcolor{darkorange}{\bf9.59}}$ & $\uline{\textcolor{royalblue}{\bf60.45}}$ & $\uline{\textcolor{royalblue}{\bf87.47}}$ & $\uline{\textcolor{royalblue}{\bf6.47}}$ & $\overline{\textcolor{darkorange}{\bf86.96}}$ & $\overline{\textcolor{darkorange}{\bf81.57}}$ & $\overline{\textcolor{darkorange}{\bf42.77}}$ & $\overline{\textcolor{darkorange}{\bf79.93}}$ & $\uline{\textcolor{royalblue}{\bf95.72}}$ & $\uline{\textcolor{royalblue}{\bf6.05}}$ \\
 &  & $\Iend/10$ & $\overline{\textcolor{darkorange}{\bf62.87}}$ & $\overline{\textcolor{darkorange}{\bf59.77}}$ & $\uline{\textcolor{royalblue}{\bf27.37}}$ & $\overline{\textcolor{darkorange}{\bf72.78}}$ & $\uline{\textcolor{royalblue}{\bf85.00}}$ & 2.73 & $\overline{\textcolor{darkorange}{\bf99.16}}$ & $\overline{\textcolor{darkorange}{\bf98.98}}$ & $\overline{\textcolor{darkorange}{\bf98.18}}$ & $\overline{\textcolor{darkorange}{\bf99.58}}$ & $\uline{\textcolor{royalblue}{\bf89.77}}$ & $\overline{\textcolor{darkorange}{\bf0.23}}$ \\
 &  & $\Iend\cdot 4/10$ & $\overline{\textcolor{darkorange}{\bf87.67}}$ & $\overline{\textcolor{darkorange}{\bf84.63}}$ & $\uline{\textcolor{royalblue}{\bf84.86}}$ & 88.33 & $\uline{\textcolor{royalblue}{\bf59.77}}$ & 0.03 & 100.00 & 100.00 & 100.00 & 100.00 & $\overline{\textcolor{darkorange}{\bf60.00}}$ & 0.00 \\
\cline{1-15} \cline{2-15}
\multirow[t]{9}{*}{Beta(0.6, 1.4)} & \multirow[t]{3}{*}{0.10} & 0 & $\overline{\textcolor{darkorange}{\bf0.19}}$ & $\uline{\textcolor{royalblue}{\bf5.03}}$ & 0.00 & $\uline{\textcolor{royalblue}{\bf0.28}}$ & $\overline{\textcolor{darkorange}{\bf65.80}}$ & $\overline{\textcolor{darkorange}{\bf4.10}}$ & $\uline{\textcolor{royalblue}{\bf2.97}}$ & $\uline{\textcolor{royalblue}{\bf8.61}}$ & $\uline{\textcolor{royalblue}{\bf0.03}}$ & $\uline{\textcolor{royalblue}{\bf20.45}}$ & $\overline{\textcolor{darkorange}{\bf78.39}}$ & $\overline{\textcolor{darkorange}{\bf5.51}}$ \\
 &  & $\Iend/10$ & $\overline{\textcolor{darkorange}{\bf0.18}}$ & $\uline{\textcolor{royalblue}{\bf5.81}}$ & $\overline{\textcolor{darkorange}{\bf0.05}}$ & $\uline{\textcolor{royalblue}{\bf0.27}}$ & $\overline{\textcolor{darkorange}{\bf65.46}}$ & $\overline{\textcolor{darkorange}{\bf2.20}}$ & $\uline{\textcolor{royalblue}{\bf18.54}}$ & $\uline{\textcolor{royalblue}{\bf18.45}}$ & $\overline{\textcolor{darkorange}{\bf10.98}}$ & $\uline{\textcolor{royalblue}{\bf26.19}}$ & $\overline{\textcolor{darkorange}{\bf76.89}}$ & $\overline{\textcolor{darkorange}{\bf1.72}}$ \\
 &  & $\Iend\cdot 4/10$ & $\overline{\textcolor{darkorange}{\bf11.73}}$ & $\uline{\textcolor{royalblue}{\bf9.02}}$ & $\uline{\textcolor{royalblue}{\bf7.30}}$ & 11.85 & $\overline{\textcolor{darkorange}{\bf54.76}}$ & 0.10 & $\overline{\textcolor{darkorange}{\bf33.05}}$ & $\uline{\textcolor{royalblue}{\bf32.03}}$ & $\overline{\textcolor{darkorange}{\bf32.63}}$ & $\uline{\textcolor{royalblue}{\bf33.32}}$ & $\overline{\textcolor{darkorange}{\bf57.88}}$ & 0.06 \\
\cline{2-15}
 & \multirow[t]{3}{*}{0.20} & 0 & $\uline{\textcolor{royalblue}{\bf3.43}}$ & $\uline{\textcolor{royalblue}{\bf10.63}}$ & $\uline{\textcolor{royalblue}{\bf0.06}}$ & $\uline{\textcolor{royalblue}{\bf6.25}}$ & $\overline{\textcolor{darkorange}{\bf77.33}}$ & $\overline{\textcolor{darkorange}{\bf6.95}}$ & $\uline{\textcolor{royalblue}{\bf15.45}}$ & $\uline{\textcolor{royalblue}{\bf23.61}}$ & $\uline{\textcolor{royalblue}{\bf0.32}}$ & $\uline{\textcolor{royalblue}{\bf52.45}}$ & $\overline{\textcolor{darkorange}{\bf90.21}}$ & $\overline{\textcolor{darkorange}{\bf7.99}}$ \\
 &  & $\Iend/10$ & $\uline{\textcolor{royalblue}{\bf4.01}}$ & $\uline{\textcolor{royalblue}{\bf14.37}}$ & $\overline{\textcolor{darkorange}{\bf1.64}}$ & $\uline{\textcolor{royalblue}{\bf8.05}}$ & $\overline{\textcolor{darkorange}{\bf76.04}}$ & $\overline{\textcolor{darkorange}{\bf3.34}}$ & $\uline{\textcolor{royalblue}{\bf57.17}}$ & $\uline{\textcolor{royalblue}{\bf56.28}}$ & $\uline{\textcolor{royalblue}{\bf43.86}}$ & $\uline{\textcolor{royalblue}{\bf67.25}}$ & $\overline{\textcolor{darkorange}{\bf86.41}}$ & $\uline{\textcolor{royalblue}{\bf1.41}}$ \\
 &  & $\Iend\cdot 4/10$ & $\overline{\textcolor{darkorange}{\bf33.11}}$ & $\uline{\textcolor{royalblue}{\bf27.84}}$ & $\overline{\textcolor{darkorange}{\bf31.73}}$ & 33.63 & $\overline{\textcolor{darkorange}{\bf57.71}}$ & 0.11 & 86.52 & $\overline{\textcolor{darkorange}{\bf85.87}}$ & $\overline{\textcolor{darkorange}{\bf86.26}}$ & $\uline{\textcolor{royalblue}{\bf86.67}}$ & $\overline{\textcolor{darkorange}{\bf59.79}}$ & 0.01 \\
\cline{2-15}
 & \multirow[t]{3}{*}{0.40} & 0 & $\uline{\textcolor{royalblue}{\bf29.19}}$ & $\uline{\textcolor{royalblue}{\bf29.37}}$ & $\uline{\textcolor{royalblue}{\bf2.17}}$ & $\uline{\textcolor{royalblue}{\bf43.28}}$ & $\overline{\textcolor{darkorange}{\bf89.58}}$ & $\overline{\textcolor{darkorange}{\bf8.20}}$ & $\uline{\textcolor{royalblue}{\bf63.83}}$ & $\uline{\textcolor{royalblue}{\bf47.62}}$ & $\uline{\textcolor{royalblue}{\bf1.81}}$ & $\uline{\textcolor{royalblue}{\bf68.89}}$ & $\overline{\textcolor{darkorange}{\bf96.65}}$ & $\overline{\textcolor{darkorange}{\bf8.56}}$ \\
 &  & $\Iend/10$ & $\uline{\textcolor{royalblue}{\bf57.91}}$ & $\uline{\textcolor{royalblue}{\bf53.26}}$ & $\overline{\textcolor{darkorange}{\bf48.79}}$ & $\uline{\textcolor{royalblue}{\bf68.30}}$ & $\overline{\textcolor{darkorange}{\bf86.18}}$ & $\uline{\textcolor{royalblue}{\bf2.69}}$ & $\uline{\textcolor{royalblue}{\bf98.99}}$ & $\uline{\textcolor{royalblue}{\bf98.71}}$ & $\uline{\textcolor{royalblue}{\bf97.44}}$ & $\uline{\textcolor{royalblue}{\bf99.36}}$ & $\overline{\textcolor{darkorange}{\bf89.83}}$ & $\uline{\textcolor{royalblue}{\bf0.19}}$ \\
 &  & $\Iend\cdot 4/10$ & $\overline{\textcolor{darkorange}{\bf87.92}}$ & $\uline{\textcolor{royalblue}{\bf84.48}}$ & $\overline{\textcolor{darkorange}{\bf87.11}}$ & 88.33 & $\overline{\textcolor{darkorange}{\bf59.78}}$ & 0.03 & 100.00 & 100.00 & 100.00 & 100.00 & $\overline{\textcolor{darkorange}{\bf60.00}}$ & 0.00 \\
\cline{1-15} \cline{2-15}
\bottomrule
\end{tabular}

    \label{tab:tab_prior_sens}\\
    MP-C: minimum power of the calibrated posterior probability of control superiority~(PPCS) test; MP-CX: minimum power of the conditional exact PPCS test; MP-UX: minimum power of the unconditional exact PPCS test; MP-W: minimum power of the asymptotic Wald test; EPASA: average expected
proportion of allocations on the superior arm.
\end{table}
	\FloatBarrier
	\section{Real-world application: ARREST trial} \label{sect:ARREST}
	In this section, we consider the effect of the
	burn-in length on the operating characteristics of the Advanced R$^2$Eperfusion STrategies for Refractory
	Cardiac Arrest~(ARREST) trial described in~\citet{yannopoulos2020advanced}, where we also consider the use of an unconditional exact test.
	The \hbox{CX-S} OST is omitted here, as the combination of this test with  early stopping is not straightforward. 
	We analyze the effect of different burn-in lengths on the unconditional OCs following from the trial design, i.e., we do not condition on the realised size of the trial. 
	
	In the ARREST trial,  {extracorporeal membrane oxygenation}~(ECMO)
	facilitated resuscitation~(developmental) was compared to
	standard advanced cardiac life support~(control) 
	in adults with an out-of-hospital
	cardiac arrest and refractory ventricular fibrillation. The inferential goal of the trial was to test~$\theta_\D=\theta_\C$ versus~$\theta_\D\geq\theta_\C$, where a success~($Y_{a,i}=1$) represented survival to hospital
	discharge.
	
	Participants were allocated treatment in groups of 30 under
	a permuted block design, with a control group allocation probability equal to the
	posterior probability (based on independent uniform priors) that the control
	treatment is superior, restricted between 0.25 and 0.75~\citep[i.e., the clip method as defined in][]{du2018comparing}. If at any of the interim analyses either the PPCS or one minus the PPCS became higher than an optional stopping threshold (OST) of~$0.986$, the recommendation was made to stop the
	trial early for futility or superiority, respectively. This OST was calibrated based on a success rate of~$0.12$ under the null hypothesis, and
	controlled type I error rate at 0.05 based on a simulation study of 10,000 samples. The considered alternative hypothesis~$\theta_\C=0.12$ and~$ \theta_\D=0.37$~(based on a treatment effect of~$0.25$) yielded a power around~$90\%$. The
	ARREST trial stopped for efficacy of the ECMO treatment after
	allocating the first group of 30 participants, with posterior probability
	of superiority of~0.9861.
	
	In~\citet{baas2025exact} this trial was re-analyzed, where the calibrated~(C) OST was compared to an UX OST. The UX OST was computed based on an extension of the Markov chain~$(\bX_t)_t$ defined in Section~\ref{sect:model}, where the extension of~$(\bX_t)_t$ also models the optional stopping component of the trial. It was shown that while the UX OST bounds type I error rate by~$0.05$ over the whole of the parameter space, it also leads to a decrease in power and EPASA. Due to the optional stopping component EPASA, assuming~$\theta_\D\geq \theta_\C$, is defined as:
	$$\mathbb{E}_\btheta^\pi[(N_{\D,i_{U}} + (\Iend-i_{U})\mathbb{I}(\text{Optional stopping in favour of~$\D$}))/\Iend]$$
	where~$U$ is the interim analysis at which the trial stops, the EPASA hence considers the fraction of participants allocated to the developmental treatment before and after optional stopping. It is hypothesised that a larger burn-in length will mitigate the large differences seen in EPASA and power across OSTs. 
	
	The ARREST trial had a maximum trial size of 150 participants and considered blocked allocation with blocks of~$30$ participants, hence we can only have~$b=15,30,45,60,75$~(noting that in the original design, the allocation probability could only be changed after~$15$ participants were allocated per arm). During the burn-in period, the target allocation proportion is equal to~$0.5$~(as before) and early stopping is only possible after allocating participant~$i\geq 2b.$ The Markov chains for computing the UX OST and the OCs described in~\citet{baas2025exact} were adapted to account for a longer burn-in length by decreasing the number of interim analyses and setting the first interim analysis at~$i=2b$. The original C OST~$0.986$ was used for~$b=15$, while for~$b>15$ the C OST was determined using the procedure outlined in~\citet[Section~5]{baas2025exact} where the only value under the null hypothesis considered was~$\theta_\C=\theta_\D=0.12$.   The same procedure~(taking all possible success rates into account) was used to determine the UX OST.
	
\begin{figure}[h!]
		\centering    
        \includegraphics[width=.92\linewidth]{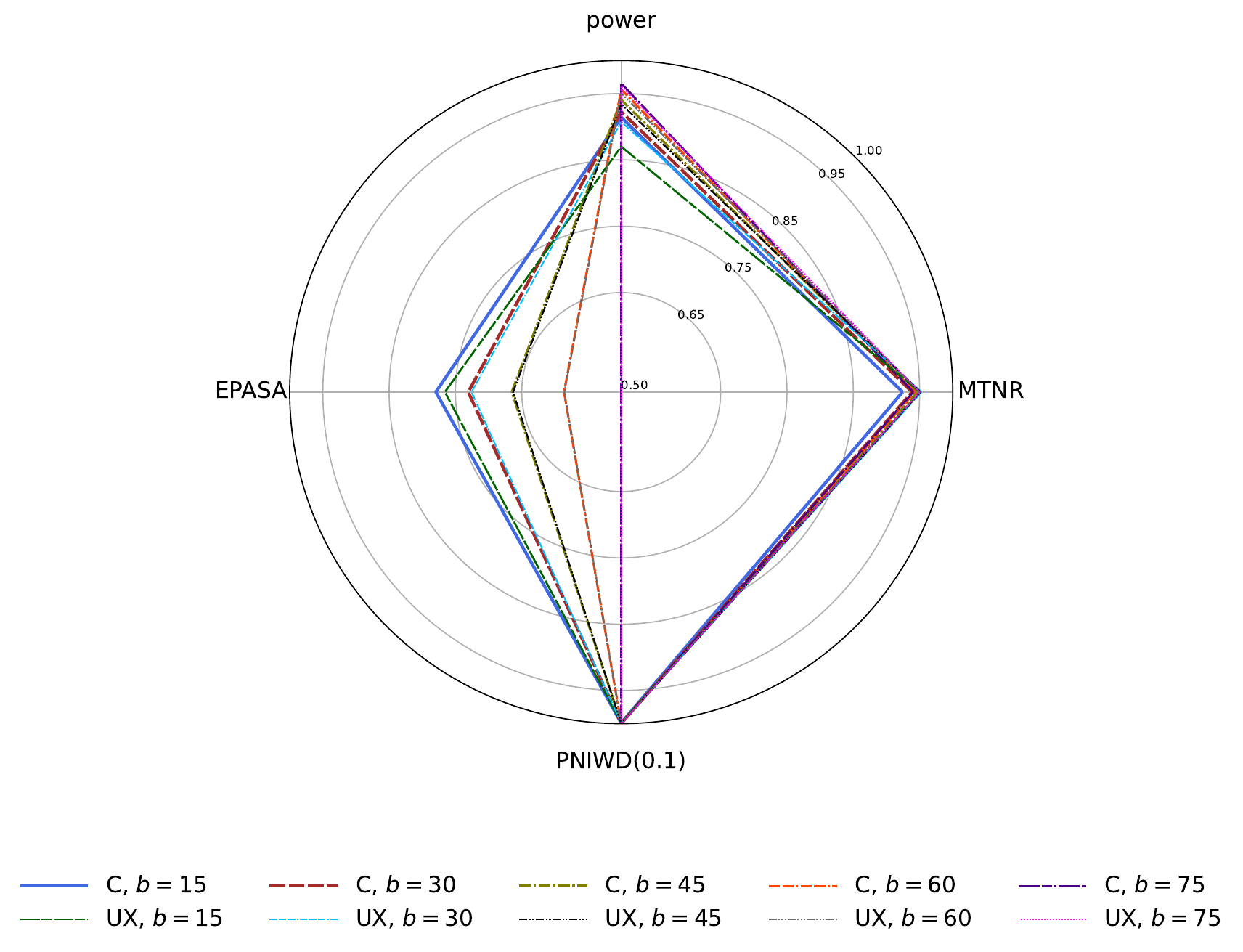}
		\caption{Star plot comparing the expected
			proportion of allocations on the superior arm (EPASA), power, minimum true negative rate over interval~$\theta_\C=\theta_\D\in\{0.02,0.03,\dots,0.22\}$~(MTNR), and probability of no imbalance in the wrong direction~(PNIWD(0.1)) for the ARREST trial when using a calibrated~(C) and unconditional exact~(UX) optional stopping threshold combined with a burn-in~$b$ of~$15,30,45,60,$ and~$75$ participants per arm.}
		\label{fig:starplot}
	\end{figure}

	Figure~\ref{fig:starplot} shows a star plot of the EPASA, power, minimum true negative rate~(MTNR, minimum value of one minus the type I error rate) and probability of no imbalance in the wrong direction~(PNIWD(0.1)). For all OCs, a higher value is better. 
	The EPASA, power, and PNIWD(0.1) were calculated under the alternative~$\theta_\C=0.12,$~$\theta_\D=0.37$, while the MTNR was calculated over the~\hbox{range~$\theta_\C=\theta_\D\in\{0.020,0.03,\dots,0.22\}$.} 
	
	Figure~\ref{fig:starplot} shows that the EPASA decreases in~$b$, while the power increases.
	The EPASA for the C OST designs is higher than for the UX OST designs due to the higher power under the C OSTs, which means that the trial stops earlier for superiority under the alternative. 
	For~$b<75,$ the type I error rate is not controlled under the C OST for~$\theta_\C=\theta_\D\in\{0.02,0.03,\dots,0.22\}$, corresponding to an MTNR less than~$0.95$, whereas for the other designs, the type I error rate is under control.  Figure~\ref{fig:starplot} shows that the PNIWD(0.1) is very similar across different designs, it ranges from~$0.999$ to~$1.00$, in agreement with Figure~\ref{fig:S01_heatmap} which shows that PIWD(0.1) is largest when treatment effects are small. As hypothesized, the differences in the OCs between the C OST and UX OST, mainly EPASA and power, decrease in~$b$.

	\FloatBarrier

	The star plot can be used to propose designs based on trade-offs in the OCs and to make a choice of clinical design, including burn-in length, based on multiple trial objectives. 
	If type I error rate control at level~$0.05$ is not of the highest importance,  Figure~\ref{fig:starplot} shows that a good option could be to use the C OST with~$b=15.$ If the type I error rate should be controlled, one design which stands out is the UX OST design with~$b=30$, showing a lower EPASA but similar power to the previously mentioned design. If EPASA is of higher importance than power, then the UX OST design with~$b=15$ might be best.  If EPASA is not important at all, the C OST design with~$b=75$ would be the best option, yielding a power higher than~$95\%$.
	
	Figure~\ref{fig:starplot_robust} shows the same evaluation 
	for a range of alternative hypotheses~$\theta_\C=\theta_\D-0.25\in\{0.02,0.03,\dots, 0.22\}$. The behaviour of the minimum PNIWD(0.1)~(MPNIWD(0.1)) in Figure~\ref{fig:starplot_robust} is similar to the PNIWD(0.1) in Figure~\ref{fig:starplot}.
	The power seems to be the most sensitive to the scenario considered, as the minimum power~(MPow) is substantially lower than in Figure~\ref{fig:starplot}.
	The MPow for the UX test with~$b=15$ is less than the (usually chosen threshold of)~$0.8$, which might be a reason not to opt for this design despite it leading to exact type I error control and high~(minimum) EPASA.

	\section{Discussion and recommendations} \label{sect:discussion}
	This paper considers the effect of an initial burn-in phase on the operating characteristics~(OCs) of the Bayesian response-adaptive randomization~(BRAR), i.e., Thompson sampling-based, design, where testing is performed either using a calibrated, conditional, or unconditional exact test. The analyses were based on the minimum, average, and maximum values of the OCs over the parameter space.

	Our numerical evaluation revealed several key insights. Firstly, calibrated or asymptotic tests exhibited significant type I error rate inflation compared to fixed designs with equal allocation, particularly in BRAR designs without a burn-in period. While increasing burn-in length offered partial mitigation, it did not eliminate the inflation entirely. Secondly, exact tests demonstrated superior power over calibration or asymptotic tests in certain parameter settings, notably with a trial size of 60 participants and varying burn-in lengths. Furthermore, the conditional exact test displayed a consistent performance profile across the parameter space and was less conservative than calibrated and asymptotic tests, even with extended burn-in periods. Third, the choice of test statistic significantly impacted statistical operating characteristics; the Wald test, for instance, yielded a more balanced type I error rate profile and greater power than tests based on the posterior probability of control superiority. Fourth, our paper considered the effect of a burn-in on estimation bias. We saw that the expected bias decreases more quickly in the burn-in length for large treatment effects and becomes higher for larger trial sizes (under the same burn-in length); hence, for controlling the bias, we advice to make the burn-in a function of the trial size, same as for the type I error rate. Our prior sensitivity analysis indicated that these finding hold across different choices of beta priors.
	
	The  evaluation in this paper uses the  approach in~\citet{baas2025exact}, where this method was applied to response-adaptive clinical trials without a  burn-in period (and mainly fully sequential trials) with up to around 1000 participants, and where it was noted that the main computational limit came from computation of the coefficients~$g^{\pi_b^B}_{i}$. Section~\ref{sect:results} states that the number of values~$ g^{\pi_b^B}_{i}$ to calculate is of order~$\mathcal{O}(\Iend^4)$ for~$b$ fixed, however the number of burn-in lengths~$b$ to consider equals~$\Iend/2+1$, making for an overall order~$\mathcal{O}(\Iend^5)$ of coefficients to compute when determining an optimal burn-in length. As~$1000^{4/5}$ is roughly 251, we expect that~$240$ participants is close to the current computational limit, however some speedups are made by parallelizing the computation of coefficients~$g^{\pi_b^B}_{i}$ over burn-ins and using the formula for~$g^{\pi^B_b}_{2b}$ given in Remark~\ref{remark:burnin}.
	
	Some general guidelines for choosing the burn-in length follow from our analysis. 
	If type I error rate control is not of the highest priority, then
	a good rule of thumb might be to choose the burn-in length per arm roughly equal to the trial size divided by four for the BRAR design. This burn-in length led to a more balanced type I error rate profile, while power was similar for higher burn-in lengths.
	In settings where type I error rate control is strictly required (e.g., in confirmatory settings), we recommend considering exact tests.
	In this case, the burn-in length could be chosen such that a sufficient power value is reached, possibly at a lower value than the trial size over four. The conditional exact test would be preferred, as our numerical evaluation showed that this test has higher power than the unconditional exact test in terms of average (and often minimum) power for small to moderate burn-in lengths. 
	
	The optimal burn-in length in terms of power or participant benefit is often
	different from the minimum or maximum possible value, hence our recommendation is to inspect more burn-in lengths than just the minimum~(i.e., zero) and the maximum (i.e., trial size over two) and inspect at least the values of~$b$ close to these two endpoints, in contrast to the guidelines given in~\citet{du2018comparing}.
	The above guidelines focus on the behaviour of the average OCs, whereas in specific settings expert opinion might better guide the burn-in length, e.g., through a desired minimum power or maximum allowed type I error rate.
	
	The penultimate section of this paper considers an illustrative application to a real-world Bayesian adaptive clinical trial with optional stopping. The trial had a complex nature, where allocation was performed in blocks based on a permuted block design and the trial stopped when, at interim, the posterior probability of superiority crossed a calibrated optional stopping threshold.  Based on the Markov chain modelling framework in~\citet{baas2025exact}, we evaluated the performance of the calibrated and unconditional optional stopping thresholds under different burn-in lengths. As it was assumed that after early stopping, the remaining participants were allocated the treatment that was deemed superior, the participant benefit depends on the optional stopping threshold of choice. The most sensitive OCs were participant benefit and power, and it was concluded that low burn-in lengths yielded a balanced trade-off. 
	
	We considered the imbalance measure, denoted    PIWD(0.1), defined by~\citet{thall2015statistical}, which quantifies the probability of substantial allocation toward the inferior arm. Consistent with existing literature, our results show that, while PIWD(0.1) increases as the absolute treatment effect decreases, the negative impact on participant benefit simultaneously diminishes. This reveals a key limitation of the measure—its dichotomous definition (imbalance vs. no imbalance) fails to account for the magnitude of the clinical effect on expected outcomes. Although fixed equal allocation minimizes PIWD(0.1), we provide a method (Section~\ref{sect:recommend}) to control PIWD(0.1) while concurrently maximizing overall participant benefit. Future research may explore alternative imbalance metrics or different parameter values.
	
	Some interesting areas are left for future research, such as an adaptive burn-in length instead of a fixed deterministic burn-in length, and the consideration of other clinical trial designs, e.g., multi-outcome or  multi-arm  designs or different outcome types.
	It is not immediately clear whether all our conclusions generalize to the latter two settings~(multiple arms and other outcome types) because these alternative settings come with their own intricacies~(e.g., in the multi-arm setting, one can perform an omnibus test, or test all arms separately, while for normal outcomes the variance is not a function of the expectation). \citet{Zhang_2020} showed type I error rate inflation under the Z-test for normal outcomes under the BRAR procedure, which we hypothesize can be mitigated but not removed completely by using a burn-in period, type I error rate control is for example guaranteed by exact tests, but which exact test performs best, and how to compute exact tests for other outcome types than binary is not established.
	
	Data collected during a burn-in period is not representative of data collected later on in the trial if there is a strong time trend. In such settings, the combination with other regularization procedures such as the clipping of allocation probabilities or a power transformation~\citep{du2018comparing} may be more appropriate, which could be explored in future research. 
	We hypothesize that, in a setting without time trends, such {\it tuning} procedures, as well as blocked allocation, will have a similar effect on overall trial OCs such as type I error rate, power, and participant benefit to using a  burn-in period, hence for each configuration of these design aspects, another burn-in length might be optimal or a burn-in might not be needed at all.
	
	
	Although we focused on BRAR designs using the posterior probability for testing, our exact evaluation method is readily applicable to analyze the effect of burn-in length across other RAR procedures \citep[e.g., those targeting optimal proportions, such as in ][where large type I error rate inflation was also observed]{pin2025revisit}. Furthermore, a comparative analysis of different calculation methods for the posterior probability of superiority would provide valuable and complementary insights to this research.
	
	While this paper addresses the underexplored issue of optimal burn-in length through numerical evaluation, underpinning these findings with theoretical results remains a key area for future research. Current theoretical literature on BRAR, largely derived from multi-armed bandit models~\citep[see, e.g.,][]{pmlr-v23-agrawal12}, often considers large-sample behavior where the effect of a finite burn-in is negligible. Furthermore, this body of work typically optimizes a single operating characteristic, contrasting with our multi-objective approach.     Further exploring the  null hypothesis behavior under BRAR~\citep[extending the work of][]{Zhang_2020} and multi-objective optimization \citep[extending the work of][]{qin2024unified} would provide valuable theoretical support for our  findings. 

\bibliographystyle{abbrvnat}

\newpage
\begin{appendices}

\begin{table}[h!]
\vspace{-6mm}
\section{Selection and description of BRAR trials with a burn-in period} \label{sect:BRARtab}
    \small
     \caption{Characteristics of multiple response-adaptive clinical trials using a burn-in period. Reference~(Ref.) number refers to the number in the  bibliography list given below.}\renewcommand{\arraystretch}{1.2}\label{tab:trial_info}
      \hspace{-7mm}
    \begin{tabular}{|p{2.5cm}|p{0.7cm}|p{0.7cm}|p{0.7cm}|p{1cm}|p{1cm}|p{1cm}|p{2cm}|p{2cm}|p{2cm}|}
        \hline
        \textbf{Trial name/number(s)} & \textbf{Ref. number}& \textbf{Phase} & \textbf{Arms} &
        \textbf{Planned burn-in period} & \textbf{Target trial size} & \textbf{Early stopping} & \textbf{Justification of burn-in period} & \textbf{Randomization rule}& \textbf{Test(s) for primary endpoint}\\
        \hline
        BATTLE / NCT00409968, NCT00411671, NCT00411632, NCT00410059, NCT00410189& 1& 2 & 4 & 80 & 250 &N & Lowest~possible to ensure $\geq1$ patient in each marker group to complete~treatment before starting RAR &  BRAR with covariate adjustment&
                 PPCS \\
        \hline
        BATTLE-2 / NCT01248247& 2
                & 2 & 4 & 70 & 200 & N & Not specified & BRAR with covariate adjustment& None\\
        \hline
        ARREST / NCT03880565&3 & 2 & 2 & 30 & $\leq$150 & Y & Not specified & BRAR with clipping &  PPCS~(early stopping threshold 0.986) \\       
        \hline
        ABT-089 / NCT00555204& 4
                & 2 & 7 & $\geq$5 per arm & $\leq$400 & Y & Not specified & BRAR &
        Posterior probability distributions/credible intervals
        \\
        \hline
        ADORE / NCT00807911 &5
                & 3 & 2 & 300 & $\leq$1100 &Y & Not specified & BRAR with variance adjustment&  PPCS~(early stopping threshold 0.99) \\
        \hline
        ESETT / NCT01960075 &6
                & 3 & 3 & 300 & $\leq$795 & Y & Not specified & BRAR & Posterior probability (interim, early stopping threshold 0.975), Chi-squared test~(final)\\
        \hline
        PAIN-CONTRoLS/ NCT02260388& 7
                & 4 & 4 & $\geq$20 per arm & $\leq$400 & Y& Not specified &BRAR & Posterior probability~(early stopping threshold 0.975)\\
        \hline
    \end{tabular}

    \label{tab:burnin}
\end{table}
\FloatBarrier

\subsection*{Bibliography (RA clinical trial list)}
\begin{enumerate}
\item Kim, E.S.,  Herbst, R. S., Wistuba, I. I., Lee, J. J., Blumenschein, G. R., Tsao, A., Stewart, D. J., Hicks, M. E., Erasmus Jr., J., Gupta, S., Alden, C. M., Liu, S., Tang, X., Khuri, F. R., Tran, H. T., Johnson, B. E., Heymach, J. V., Mao, L., Fossella, F., Kies, M. S., Papadimitrakopoulou, V., Davis, S. E., Lippman, S. M., Hong, W. K. (2011). {\it The BATTLE trial: Personalizing therapy for lung cancer. Cancer Discovery 1}(1), 44–53. \url{https://doi.org/10.1158/2159-8274.CD-10-0010}
\item Papadimitrakopoulou, V., Lee, J. J., Wistuba, I. I., Tsao, A. S., Fossella, F. V., Kalhor, N., Gupta, S., Byers, L. A., Izzo, J. G., Tang, X., Skoulidis, F., Gibbons, D. L., Shen, L., Wei, C., Diao, L., Peng, S. A., Wang, J., Tam, A. L., Heymach, J. V., Hong, W. K., Gettinger, S. N., Goldberg, S. B., Koo, J. S., Herbst, R. S., Miller, V. A., Coombes, K. R., Mauro, D. J., \& Rubin, E. H. (2016). The BATTLE-2 study: A biomarker-integrated targeted therapy study in previously treated patients with advanced non-small-cell lung cancer. {\it Journal of Clinical Oncology, 34}(30), 3638–3647. \url{https://doi.org/10.1200/JCO.2015.66.0084}
\item D. Yannopoulos, J. Bartos, G. Raveendran, E. Walser, J. Connett, T. A. Murray, G. Collins, L. Zhang, R. Kalra,
M. Kosmopoulos, R. John, A. Shaffer, R. J. Frascone, K. Wesley, M. Conterato, M. Biros, J. Tolar, and T. P.
Aufderheide. Advanced reperfusion strategies for patients with out-of-hospital cardiac arrest and refractory
ventricular fibrillation (ARREST): A phase 2, single centre, open-label, randomised controlled trial. {\it The Lancet,
396}(10265):1807–1816, 2020.  \url{https://doi.org/10.1016/S0140-6736(20)32338-2}.
\item Lenz, R. A., Pritchett, Y. L., Berry, S. M., Llano, D. A., Han, S., Berry, D. A., Sadowsky, C. H., Abi-Saab, W. M., \& Saltarelli, M. D. (2015). Adaptive, dose-finding phase 2 trial evaluating the safety and efficacy of ABT-089 in mild to moderate Alzheimer disease. {\it Alzheimer Disease \& Associated Disorders, 29}(3), 192–199. \url{https://doi.org/10.1097/WAD.0000000000000093}
\item Carlson, S. E., Gajewski, B. J., Valentine, C. J., Kerling, E. H., Weiner, C. P., Cackovic, M., Buhimschi, C. S., Rogers, L. K., Sands, S. A., Brown, A. R., Mudaranthakam, D. P., Crawford, S. A., \& DeFranco, E. A. (2021). Higher dose docosahexaenoic acid supplementation during pregnancy and early preterm birth: A randomised, double-blind, adaptive-design superiority trial. {\it EClinicalMedicine, 36}, 100905. \url{https://doi.org/10.1016/j.eclinm.2021.100905 }
 \item Chamberlain, J. M., Kapur, J., Shinnar, S., Elm, J., Holsti, M., Babcock, L., Rogers, A., Barsan, W., Cloyd, J., Lowenstein, D., Bleck, T. P., Conwit, R., Meinzer, C., Cock, H., Fountain, N. B., Underwood, E., Connor, J. T., Silbergleit, R., Neurological Emergencies Treatment Trials, \& Pediatric Emergency Care Applied Research Network investigators. (2020). Efficacy of levetiracetam, fosphenytoin, and valproate for established status epilepticus by age group (ESETT): A double-blind, responsive-adaptive, randomised controlled trial. {\it The Lancet, 395}(10231), 1217–1224. \url{https://doi.org/10.1016/S0140-6736(20)30611-5}
\item  Barohn, R. J., Gajewski, B., Pasnoor, M., Brown, A., Herbelin, L. L., Kimminau, K. S., Mudaranthakam, D. P., Jawdat, O., Dimachkie, M. M., \& the PAIN-CONTRoLS Study Team. (2020). Patient Assisted Intervention for Neuropathy: Comparison of Treatment in Real Life Situations (PAIN-CONTRoLS): Bayesian adaptive comparative effectiveness randomized trial. {\it JAMA Neurology, 77}(5), 557–567. \url{https://doi.org/10.1001/jamaneurol.2020.2590}
\end{enumerate}

\begin{figure}[h!]
\section{Type I error rate and power results for the Wald test}\label{RR_wald}
    \centering
    \includegraphics[width=\linewidth]{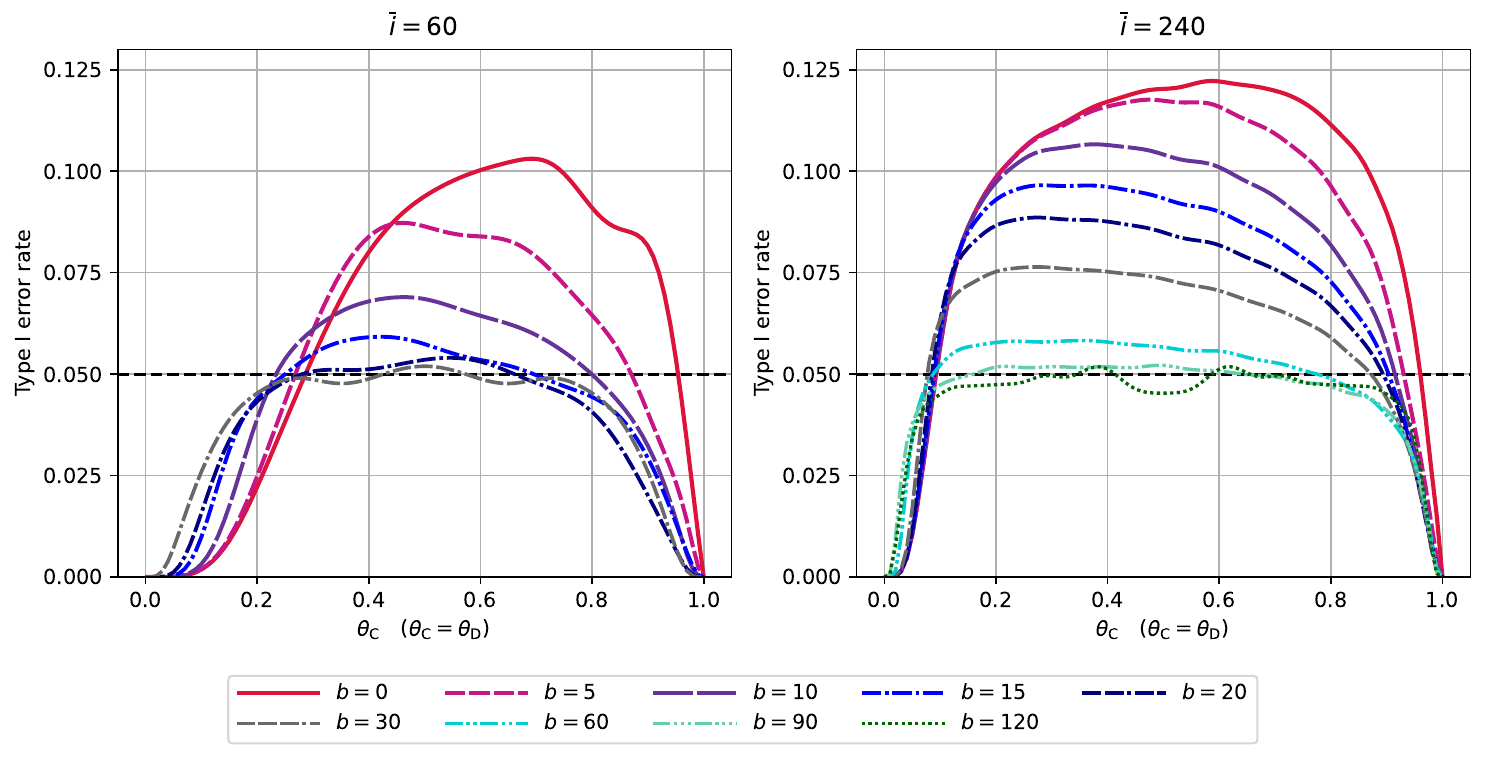}
    \caption{Type I error rate profiles for the Bayesian response-adaptive randomization design using the asymptotic Wald test for trial sizes~$\bar{i}=60$ and~$\bar{i}=240$, across~$\theta_\C=\theta_\D\in \{0.00,0.01,\dots,1.00\}$ for different values of burn-in length~$b$. The significance level~$\alpha$ was set to~$0.05$.}
    \label{fig:AS_N=60,240}
\end{figure}
\begin{table}[h!]
\centering
\small
\caption{Maximum type I error rate across the whole parameter space~(using a grid approximation) (in \%) of the asymptotic Wald test based on the posterior probability of control superiority~(significance level~$\alpha=0.05$) under the Bayesian response-adaptive randomization design as burn-in varies. Burn-in proportion (BP) is the proportion of the trial within the burn-in stage, given by BP$=2b/\bar{i}$. BP$=1.0$ corresponds to equal allocation. Type I error rates above 6\% are indicated in bold, with the severity of the inflation emphasized by a gradient (red being worst).}\label{tab:maxAStype1}
\begin{tabular}{lccccccccccc}
\toprule
BP& 0.00 & 0.10 & 0.20 & 0.30& 0.40 & 0.50 & 0.60& 0.70& 0.80 & 0.90 & 1.00 \\
\midrule
$\Iend = 20$ & \textcolor{brown}{\bf 6.04} & \textcolor{brown}{\bf 6.22} & \textcolor{brown}{\bf 7.05} & \textcolor{brown}{\bf 6.75} & \textcolor{brown}{\bf 
6.09} & 5.34 & 4.76 & 5.90 & 5.53 & 5.12 & 4.22 \\
$\Iend = 40$ & \textcolor{orange}{\bf 9.72} & \textcolor{orange}{\bf 9.89} & \textcolor{brown}{\bf 7.88} & \textcolor{brown}{\bf 7.00} & \textcolor{brown}{\bf 
6.30} & \textcolor{brown}{\bf 6.08} & 5.65 & 5.94 & 5.69 & 5.16 & 5.19 \\
$\Iend = 60$ & \textcolor{red}{\bf 10.31} & \textcolor{orange}{\bf 9.87} & \textcolor{orange}{\bf 8.21} & \textcolor{brown}{\bf 7.11} & \textcolor{brown}{\bf 6.54} & 5.92 & 5.52 & 5.16 & 5.17 & 5.20 & 5.19 \\
$\Iend = 80$ & \textcolor{red}{\bf 10.86} & \textcolor{red}{\bf 10.07} & \textcolor{orange}{\bf 8.24} & \textcolor{brown}{\bf 7.07} & \textcolor{brown}{\bf 6.28} & 5.81 & 5.47 & 5.32 & 5.09 & 5.08 & 5.67 \\
$\Iend = 100$ & \textcolor{red}{\bf 11.27} & \textcolor{red}{\bf 10.17} & \textcolor{orange}{\bf 8.25} & \textcolor{brown}{\bf 7.10} & \textcolor{brown}{\bf 6.30} & 5.82 & 5.57 & 5.29 & 5.23 & 5.22 & 5.69 \\
$\Iend = 240$ & \textcolor{red}{\bf 12.22} & \textcolor{red}{\bf 10.23} & \textcolor{orange}{\bf 8.32} & \textcolor{brown}{\bf 7.10} & \textcolor{brown}{\bf 6.35} & 5.83 & 5.54 & 5.26 & 5.11 & 5.10 & 5.18 \\
\bottomrule
\end{tabular}\end{table}

\FloatBarrier

\begin{figure}
    \centering
    \includegraphics[width=\linewidth]{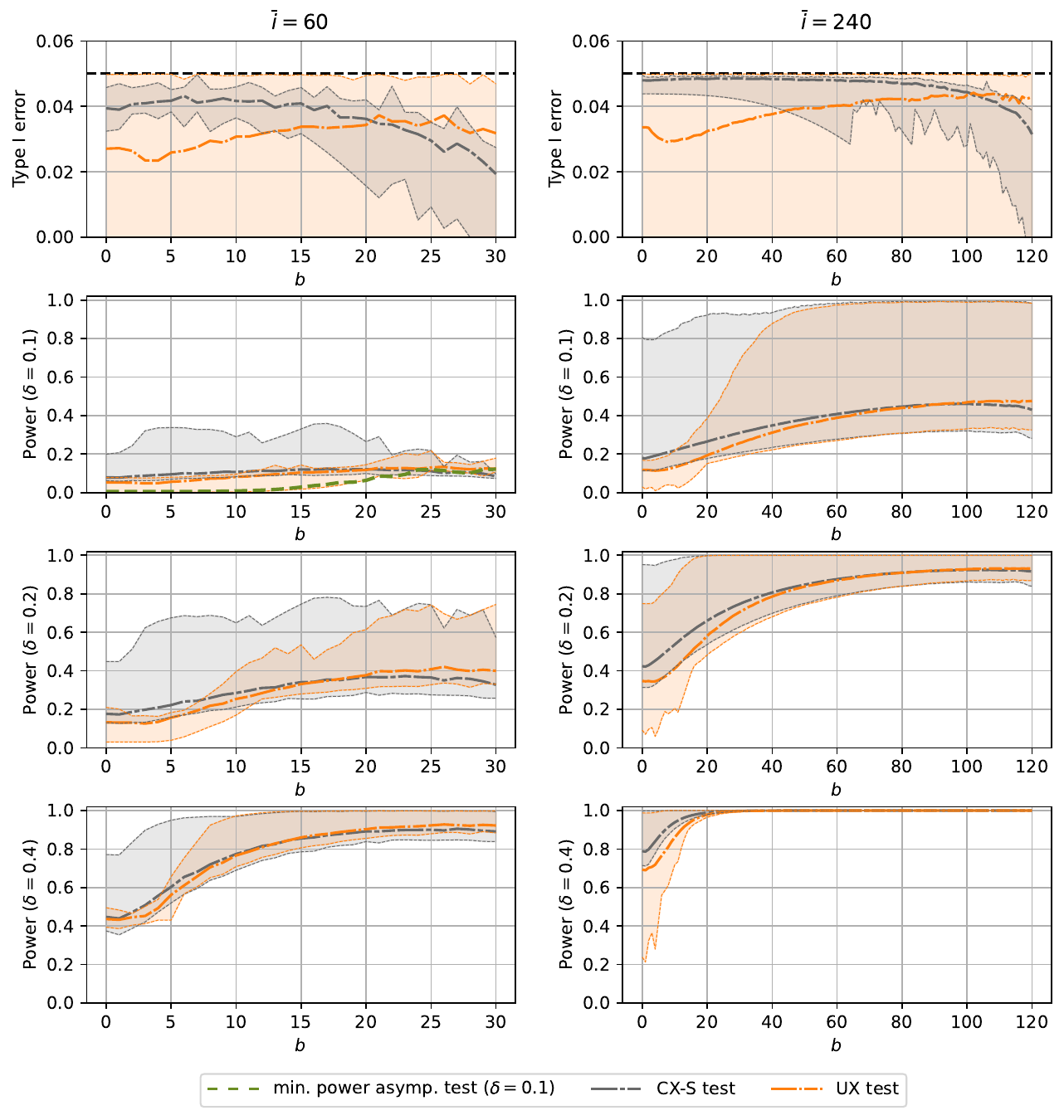}
    \caption{Type I error rate and power under detect treatment effects~$\delta\in \{0.1,0.2,0.4\}$ for the Bayesian response-adaptive randomization design using the unconditional exact~(UX) and conditional exact Wald test based on total successes (CX-S  test) with trial size~$\Iend=60$ and~$\Iend=240$ across different burn-in lengths. For comparison we have plotted the (grid-approximated) minimum power for the asymptotic Wald test for~$\Iend=60$ and~$\delta=0.1$.
         The averages for each treatment effect~$\delta$ and fixed burn-in length are represented by thick lines, while the minimum and maximum are presented by thin dotted lines and the ribbons.  The significance level~$\alpha$ was set to~$0.05$.}
    \label{fig:typeIall}
\end{figure}
\begin{figure}
    \centering
    \includegraphics[width=.65\linewidth]{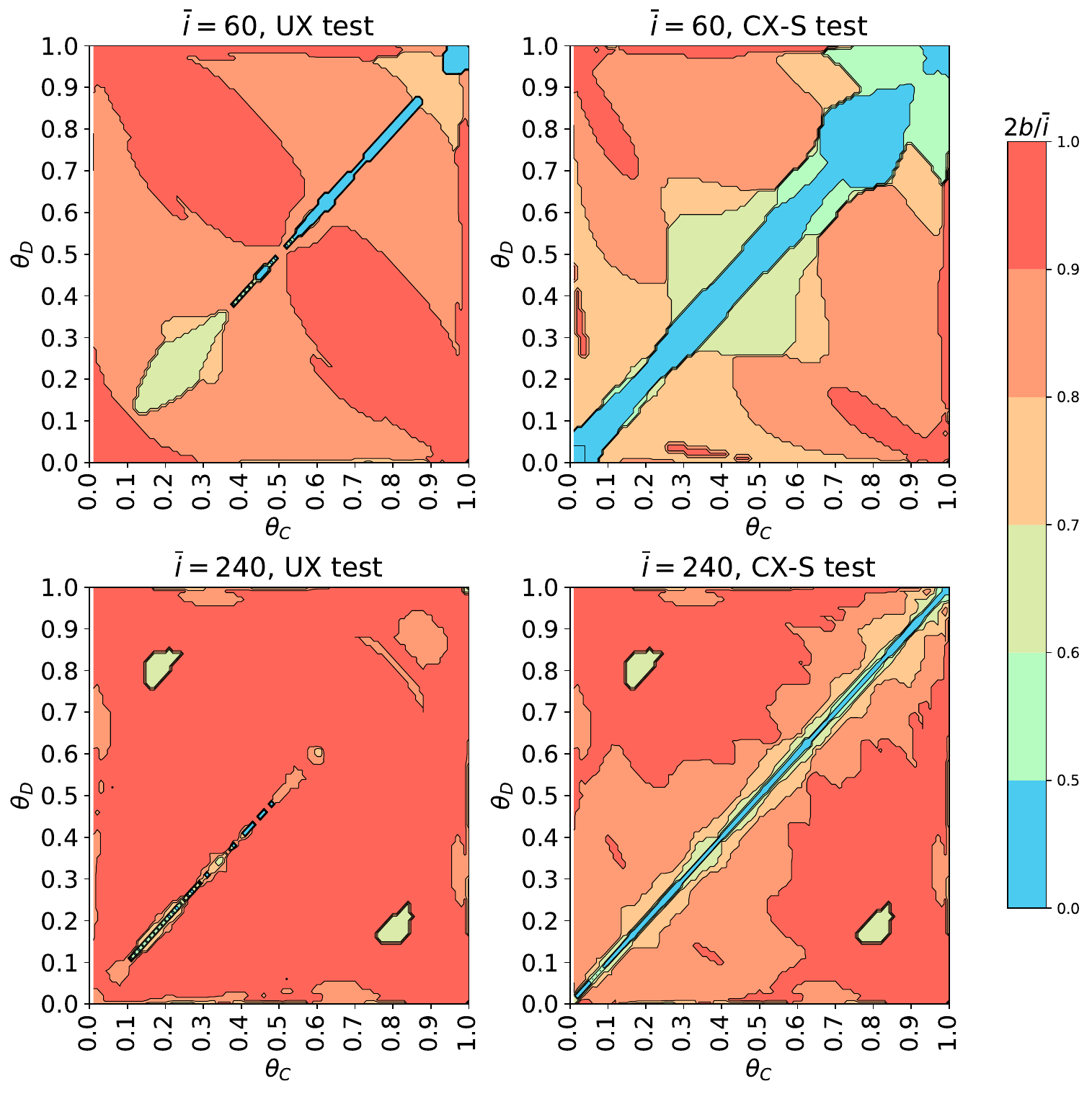}
    \caption{Optimal  burn-in proportions~($2b/\Iend$) in terms of power of the Bayesian response-adaptive randomization design using the unconditional exact~(UX) and conditional exact test based on total successes~(CX-S test) based on the Wald test statistic for every parameter configuration 
in the grid~$\theta_\C,\theta_\D\in\{0.00,0.01,\dots, 1.00\}.$ The considered trial sizes are~$\Iend=60$ and~$\Iend=240$ participants.}
    \label{fig:optimal_burnin_Wald}
\end{figure}
\FloatBarrier

\begin{table}[h!]
\section{Tables with critical values}~\label{app:critvals}
\label{table:crit}
\setlength{\tabcolsep}{3pt}
\caption{Upper critical values per burn-in length~$b$ for trial size~$\Iend=60$ for the calibration test based on the posterior probability of control superiority~(PPCS), the unconditional exact~(UX) test based on PPCS, conditional exact PPCS test based on total successes (CX-S PPCS test) for 
12 and 48 total successes, as well as the upper critical values when using the Wald test statistic for these tests instead of the PPCS. The right significance level was set to~$2.5\%$. Trailing zeros are not displayed.}
\hspace{-3mm}
\scriptsize
\begin{tabular}{clllllll}\hline
b & Calibration (PPCS) & UX PPCS & CX-S PPCS (S=12) & CX-S PPCS (S=48) & UX Wald & CX-S Wald (S=12) & CX-S Wald (S=48)\\\hline
0 & 0.978233355395697 & 0.9947496072990138 & 0.9485264395008914 & 0.9927984157010034 & 2.302718174896149 & 3.9735970711951314 & 16.685322891691367\\
1 & 0.9781225779090684 & 0.9953212845230767 & 0.9485264395008914 & 0.9927984157010034 & 2.302718174896149 & 3.9735970711951314 & 16.685322891691367\\
2 & 0.978424114225277 & 0.992067689053411 & 0.9485264395008914 & 0.9927984157010034 & 2.314198293489495 & 3.9735970711951314 & 2.643304042534928\\ 
3 & 0.9790260525484195 & 0.9916672873845506 & 0.9485264395008914 & 0.9861681852935872 & 2.3348799378810092 & 3.9735970711951314 & 1.969777651208623\\
4 & 0.9805847153830307 & 0.9916672873845501 & 0.9485264395008914 & 0.9861681852935872 & 2.3069908920018367 & 3.9735970711951314 & 2.0000000000000004\\
5 & 0.9805847153830307 & 0.9898612626802341 & 0.9524953443194414 & 0.9861681852935872 & 2.244193077610112 & 3.9735970711951314 & 1.9560942740767506\\
6 & 0.979483984434495 & 0.9884376484509019 & 0.957389825103261 & 0.9850237556333951 & 2.217970823170696 & 3.9735970711951314 & 1.9103280233981044\\7 & 0.9790260525484195 & 0.9881797899504889 & 0.957389825103261 & 0.9850237556333951 & 2.1920662875450887 & 3.986942522942203 & 1.894212202238607\\8 & 0.9786845650880073 & 0.9861019884949906 & 0.957389825103261 & 0.9850237556333951 & 2.1496104640952374 & 3.986942522942203 & 1.894212202238607\\9 & 0.9786845650880073 & 0.9849109348028043 & 0.957389825103261 & 0.9794839844346975 & 2.146561442983005 & 3.986942522942203 & 1.9420166249104487\\10 & 0.9784545138738479 & 0.9848017659048213 & 0.9608623622030332 & 0.9848973252369821 & 2.1145598486192863 & 3.986942522942203 & 2.0000000000000004\\
11 & 0.9781640004387606 & 0.9839067305268394 & 0.969962204991043 & 0.9794839844346975 & 2.0961796002305153 & 4.014260294784796 & 1.9103280233981044\\
12 & 0.9776156670306634 & 0.9839067305268394 & 0.969962204991043 & 0.9794839844346975 & 2.083022575693182 & 4.014260294784796 & 1.830016782527511\\13 & 0.9764862727469846 & 0.9820200951877319 & 0.9768885222819681 & 0.9779145861440016 & 2.0660294406384994 & 4.029304421069107 & 1.9103280233981044\\
14 & 0.9764862727469846 & 0.9820200951877319 & 0.9663733717955911 & 0.9848973252369821 & 2.06568306450296 & 2.388037076469809 & 1.969777651208623\\15 & 0.9763391532922311 & 0.9818031890578925 & 0.9608623622030332 & 0.9779145861440016 & 2.021702176265099 & 2.388037076469809 & 1.9103280233981044\\
16 & 0.9759825519641755 & 0.9800594713363734 & 0.9608623622030332 & 0.9775738544502989 & 2.021702176265099 & 2.388037076469809 & 1.830016782527511\\
17 & 0.9757243012088005 & 0.9797834542199031 & 0.9608623622030332 & 0.9779145861440016 & 2.0217021762650984 & 2.388037076469809 & 1.9103280233981044\\
18 & 0.9756621681019739 & 0.9792993282725425 & 0.9707139671387696 & 0.9847392002468287 & 2.011594300028457 & 2.5131234497501733 & 2.0283702113484394\\
19 & 0.9759825519641755 & 0.9775738544503001 & 0.9707139671387696 & 0.9779145861440016 & 2.011594300028457 & 2.5131234497501733 & 1.9103280233981044\\
20 & 0.9761502054406046 & 0.9775738544503001 & 0.9784545138739889 & 0.9779145861440016 & 2.011594300028457 & 2.635081948268323 & 1.9103280233981044\\
21 & 0.9749982908819077 & 0.9774557869401455 & 0.9726534336515967 & 0.9683933684856809 & 1.9639359620800905 & 2.227784021228921 & 1.7747405374280212\\
22 & 0.9749982908819077 & 0.975475696405277 & 0.9663733717955911 & 0.9830678759741764 & 1.9924984760531164 & 2.1064127776663257 & 2.078036923180232\\
23 & 0.9749982908819077 & 0.975235086733878 & 0.9663733717955911 & 0.9830678759741764 & 1.9924984760531164 & 2.1064127776663257 & 2.078036923180232\\
24 & 0.9761854762466753 & 0.9766456353512373 & 0.9663733717955911 & 0.9830678759741764 & 2.0013843300632175 & 2.1064127776663257 & 2.078036923180232\\
25 & 0.9742791605266793 & 0.9754760637149245 & 0.9726534336515967 & 0.9734846398942882 & 1.9638370552058695 & 2.227784021228921 & 1.9420166249104487\\
26 & 0.9736205692491952 & 0.97401977359379 & 0.9803217074512501 & 0.9631244804017428 & 1.9493917067918949 & 2.3397979352670997 & 1.79867847541553\\27 & 0.9761854762466762 & 0.9761854762466762 & 0.9852683841251674 & 0.9515529995231619 & 1.9936203984193257 & 2.4249119918964435 & 1.6753366866562536\\
28 & 0.976429464794035 & 0.9769204877646053 & 0.9628077487730287 & 0.9845660747738446 & 2.010841003245581 & 1.8841275935760935 & 2.2472576881857393\\
29 & 0.9785993142093347 & 0.9785993142093347 & 0.9628077487730287 & 0.9790555184650054 & 2.0455920583489493 & 1.8841275935760935 & 2.1467914292930836\\
30 & 0.9793538324121724 & 0.9793538324121724 & 0.9723027995475091 & 0.9723027995475023 & 2.06568306450296 & 2.0219862629219842 & 2.0219862629219842\\\hline
\end{tabular}

\end{table}

\begin{table}[h!]

\setlength{\tabcolsep}{3pt}
    \caption{Upper critical values per burn-in length~$b$ for trial size~$\Iend=240$ for the calibration test based on the posterior probability of control superiority~(PPCS), the unconditional exact~(UX) test based on PPCS, conditional exact PPCS test based on total successes (CX-S PPCS test) for 24 and 96 total successes, as well as the upper critical values when using the Wald test statistic for these tests instead of the PPCS. The right significance level was set to~$2.5\%$. Trailing zeros are not displayed.}
    \label{tab:crit240}
    \hspace{-3mm}
    \scriptsize
\begin{tabular}{clllllll}\hline
b & Calibration (PPCS) & UX PPCS & CX-S PPCS (S=24) & CX-S PPCS (S=96) & UX Wald & CX-S Wald (S=24) & CX-S Wald (S=96)\\\hline
0 & 0.9882501953326069 & 0.9978695860910114 & 0.9935673601041999 & 2.4924269802206607 & 0.976712301838966 & 3.423408216346033 & 2.2779791898059987\\
3 & 0.9882781171698278 & 0.9966560452481168 & 0.9935673601041999 & 2.4911702387121983 & 0.976712301838966 & 3.423408216346033 & 2.1738595113117185\\
6 & 0.9879776837660381 & 0.9949247272897915 & 0.991084274255219 & 2.4685982065131458 & 0.9779237683765297 & 3.4027103916098396 & 2.066154702523348\\
9 & 0.9868309816266577 & 0.994149816164298 & 0.9899618030235767 & 2.424635510269878 & 0.9779237683765297 & 3.4027103916098396 & 2.010821304062485\\12 & 0.9858978257750575 & 0.9931308790640347 & 0.9899618030235767 & 2.3774454637273075 & 0.9795839919025272 & 3.315330264217996 & 1.9892287793072134\\
15 & 0.9848564964812686 & 0.9922694364958993 & 0.9883302461014646 & 2.3342462996474693 & 0.9795839919025272 & 3.248972771446206 & 1.9879588372011827\\
18 & 0.9840517447847466 & 0.991454930327456 & 0.9875684653705998 & 2.297411946432368 & 0.9789473370146099 & 3.0786169791242903 & 1.9724803123384034\\
21 & 0.9834677828659331 & 0.99040429280052 & 0.9863823116211614 & 2.2540856718479256 & 0.9779237683765297 & 2.836017769632135 & 1.9724803123384034\\
24 & 0.9824742885518879 & 0.989389294521634 & 0.9851695377108294 & 2.225770355248036 & 0.9779237683765297 & 2.7329570802983096 & 1.92740863486254\\27 & 0.981972612621218 & 0.9890253991729132 & 0.9843824508623696 & 2.1972067198446776 & 0.9779237683765297 & 2.711737464386029 & 1.9242014021264486\\
30 & 0.9813702981129963 & 0.988098859326659 & 0.9841871719052863 & 2.173363798231878 & 0.976712301838966 & 2.628521839387073 & 1.9193746913530028\\33 & 0.9806907493468144 & 0.987111560547467 & 0.9838671977559389 & 2.1505817024682106 & 0.9759029039046823 & 2.5396662053860934 & 1.9059541995122709\\
36 & 0.9801766726149658 & 0.9860142943329693 & 0.9837443758363629 & 2.132766488487511 & 0.9761420901175495 & 2.4538755828982217 & 1.9059541995122709\\
39 & 0.9798407915291214 & 0.9859446520122928 & 0.983429964637266 & 2.1152174018997143 & 0.9759029039046823 & 2.4504851191518493 & 1.9059541995122709\\
42 & 0.9792761345235556 & 0.985801486990717 & 0.9813391438152439 & 2.102010609415397 & 0.9759029039046823 & 2.4221470462812964 & 1.9242014021264486\\
45 & 0.9788214804461866 & 0.9847863771303834 & 0.9813391438152439 & 2.082607348876787 & 0.9757560212672464 & 2.336610829863881 & 1.902062854834968\\
48 & 0.97826537183464 & 0.984614262033002 & 0.9812539859241819 & 2.0717534025881643 & 0.9757560212672464 & 2.336610829863881 & 1.902062854834968\\ 
51 & 0.9780269497861982 & 0.9838908810985367 & 0.9813391438152439 & 2.06247788425256 & 0.9757560212672464 & 2.2648211721355853 & 1.902062854834968\\
54 & 0.9777912401878406 & 0.9829776928095112 & 0.979841597065226 & 2.0571789450556857 & 0.9757560212672464 & 2.2648211721355853 & 1.899702886341581\\
57 & 0.9775302625937538 & 0.9821405301601455 & 0.979841597065226 & 2.0457693746884837 & 0.9729880772227908 & 2.2195169446627285 & 1.899702886341581\\
60 & 0.9771025081368233 & 0.9813949578663805 & 0.979841597065226 & 2.0385605822215553 & 0.9729880772227908 & 2.2195169446627285 & 1.899702886341581\\
63 & 0.9769494992137495 & 0.9802557008852985 & 0.9792028055195411 & 2.034581972198286 & 0.9729880772227908 & 2.194420017872711 & 1.9193746913530028\\
66 & 0.9766889592173608 & 0.9800704486354284 & 0.9792028055195411 & 2.022816583553441 & 0.9729880772227908 & 2.1856111388590524 & 1.9193746913530028\\
69 & 0.976389370445761 & 0.9795346029830811 & 0.9792028055195411 & 2.01261895935118 & 0.9729055089671458 & 2.1104700481554057 & 1.9193746913530028\\
72 & 0.9761236193534458 & 0.9789083257637078 & 0.9773763117006349 & 2.0091250451709657 & 0.9745236227530435 & 2.18049550041504 & 1.896246417718264\\
75 & 0.9759873910276765 & 0.9803586199033842 & 0.9773763117006349 & 2.0014074321358986 & 0.9722155756458432 & 2.1104700481554057 & 1.896246417718264\\
78 & 0.9757055519544848 & 0.9798372471904924 & 0.9793750452494494 & 1.9996422180924787 & 0.9744379445372018 & 2.118499646075263 & 1.9468942627400265\\
81 & 0.9756273686052352 & 0.9783026447072881 & 0.9780120571464848 & 1.9996422180924787 & 0.9729055089671458 & 2.0689765493569414 & 1.9378684892556755\\
84 & 0.9754874136349871 & 0.9774463867226041 & 0.9780120571464848 & 1.9958109889528413 & 0.9744379445372018 & 2.118499646075263 & 1.9378684892556755\\
87 & 0.9753200810730491 & 0.9771599995279764 & 0.9766902783228439 & 1.9848970621076392 & 0.9729055089671458 & 2.0689765493569414 & 1.92740863486254\\
90 & 0.975182270366414 & 0.9783167736432662 & 0.9755905293485424 & 1.9807336583453024 & 0.9766494332233285 & 2.1400789498594053 & 1.9231804685955098\\
93 & 0.9753080693581938 & 0.9761808390664579 & 0.9755905293485424 & 1.979380805888517 & 0.9716880143896327 & 2.024206425597235 & 1.9224770137746274\\
96 & 0.9751324173952217 & 0.9766414148640596 & 0.9783868122902123 & 1.9803981004567102 & 0.979156942716413 & 2.1540449283661682 & 1.9842047686233453\\
99 & 0.9750008632856821 & 0.9756726757511633 & 0.9748917874938483 & 1.9812003226878416 & 0.9746109508557322 & 2.0501494047420614 & 1.9209360365088135\\
102 & 0.9750317553371384 & 0.9768872972234418 & 0.9779137701086565 & 1.9706219415562913 & 0.9738391142340772 & 2.022053897915093 & 1.9879261792733791\\
105 & 0.9749245867691738 & 0.9756342185057608 & 0.9739593959105642 & 1.972333132493384 & 0.9738391142340772 & 2.022053897915093 & 1.9231804685955098\\
108 & 0.9745878486997613 & 0.9751764047282466 & 0.9772698848404712 & 1.970246478514981 & 0.9733655062190011 & 1.9980807374984912 & 1.9912645079497961\\
111 & 0.9747306272212816 & 0.9753831265761124 & 0.9730417715937933 & 1.9902213647212097 & 0.977083942124909 & 2.060251454007429 & 1.9398677154646626\\
114 & 0.9745692173455381 & 0.9754534449717089 & 0.9730417715937933 & 1.9710902033222655 & 0.9730326441004481 & 1.9696214188988213 & 1.9398677154646626\\
117 & 0.9741029707535324 & 0.9753250826946339 & 0.9767689265551889 & 1.9736254009318919 & 0.9802003247807616 & 2.102342700918535 & 2.0179640311794094\\
120 & 0.9734159553522349 & 0.9753797363964998 & 0.9729459236390567 & 1.9711384650967143 & 0.9729459236386523 & 1.952913461478469 & 1.9529134614784684\\\hline
\end{tabular}
\end{table}

\begin{figure}[h!]
\section{Additional results ARREST trial analysis}
    \centering
    \includegraphics[width=.92\linewidth]{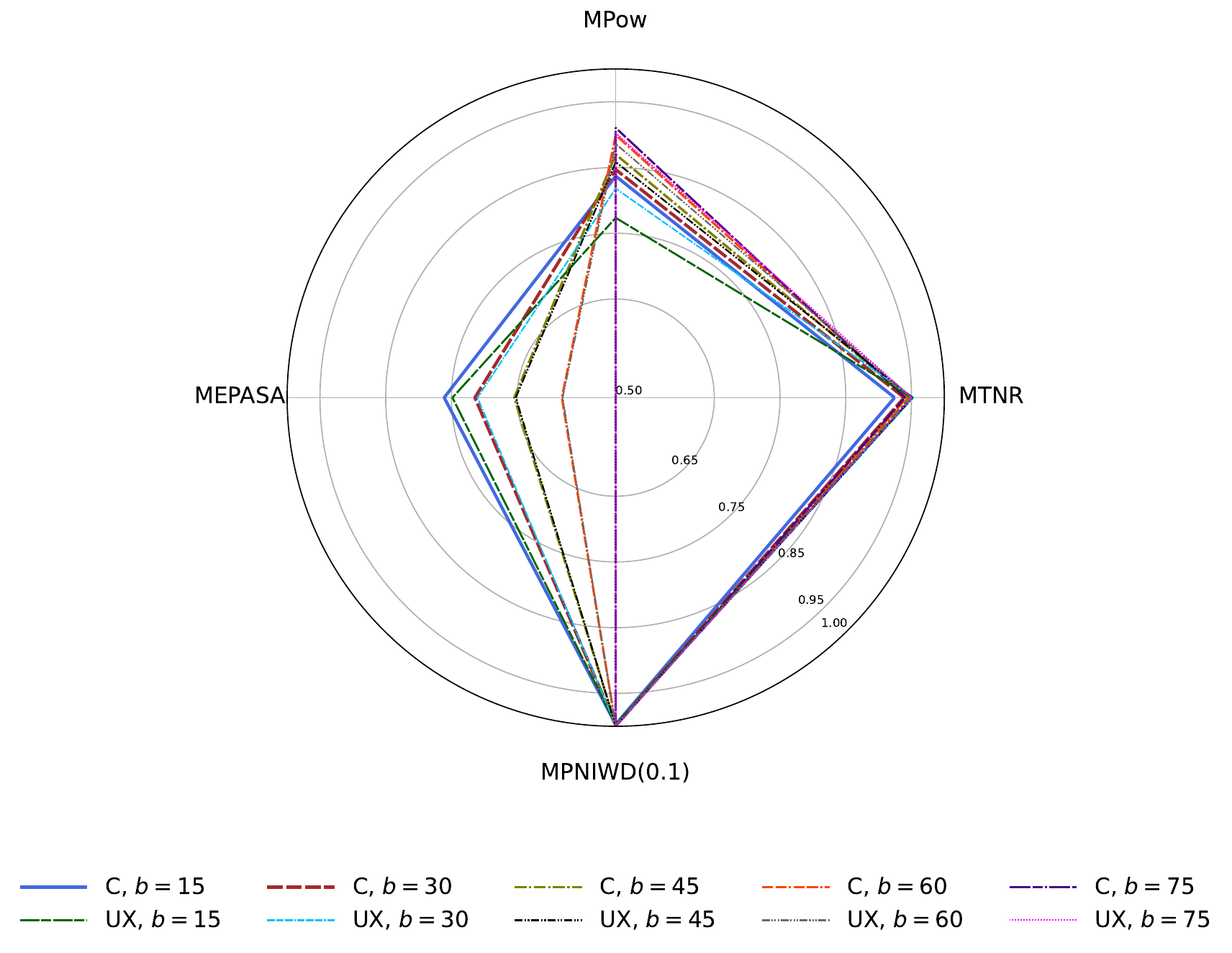}
    \caption{Star plot comparing the minimum expected
proportion of allocations on the superior arm~(MEPASA), minimum power~(MPow), minimum true negative rate over interval~$\theta_\C=\theta_\D\in\{0.02,0.03,\dots,0.22\}$~(MTNR), and minimum probability of no imbalance in the wrong direction~(MPNIWD(0.1)) for the ARREST trial when using a calibrated~(C) and unconditional exact~(UX) optional stopping threshold combined with a burn-in~$b$ of~$15,30,45,60,$ and~$75$ participants per arm. The minimum is taken over~$\theta_\C=\theta_\D-\delta\in\{0.02,0.03,\dots, 0.22\}$ where~$\delta=0$ for the null hypothesis and~$\delta=0.25$ for the alternative hypothesis.}
    \label{fig:starplot_robust}
\end{figure}

\end{appendices}
\end{document}